\documentclass[final,5p,times,twocolumn]{elsarticle}

\usepackage{amsmath}
\usepackage{amssymb}
\usepackage{bm}
\usepackage{mathrsfs}

\usepackage{graphicx}
\usepackage{lscape}
\usepackage{epstopdf}
\usepackage{subfigure}
\usepackage{bmpsize}

\usepackage{natbib}
\usepackage{pxfonts}

\usepackage{xr}
\usepackage{color}

\usepackage{multirow}
\usepackage{booktabs}  
\usepackage{threeparttable}

\usepackage{fancyhdr}
\usepackage[hang,flushmargin]{footmisc}

\usepackage[colorlinks, linkcolor=blue, anchorcolor=blue, citecolor=blue]{hyperref}

\journal{Acta Materialia}

\begin{document}
	\begin{frontmatter}
		
		\title{\textbf{A practical coarse-grained formula for classical mobility of interstitial helium diffusion in BCC W and Fe}}
		
		\author[label_IFCEN,label_PhysDep]{Haohua Wen}
		\author[label_IFCEN,label_PhysDep]{Jianyi Liu}
		\author[label_PhysDep]{Yifeng Wu}
		\author[label_IFCEN,label_PhysDep]{Kan Lai}
		
		\author[label_PhysDep,label_KeyLab]{Yue Zheng\corref{CA1}}
		\cortext[CA1]{Corresponding author: zhengy35@mail.sysu.edu.cn}
		
		\address[label_IFCEN]{
			Sino-French Institute of Nuclear Engineering and Technology, Sun Yat-sen University, Zhuhai 519082, China
		}
	
		\address[label_PhysDep]{
			Micro\&Nano Physics and Mechanics
			Research Laboratory, School of Physics, Sun Yat-sen University, Guangzhou 510275, China
		}
		
		\address[label_KeyLab]{
			State Key Laboratory of Optoelectronic Materials and Technologies, School of Physics, Sun Yat-sen University, Guangzhou 510275, China
		}
	
\begin{abstract}
	Helium diffusion in metals is the basic requirement of nucleation and growth of bubble, which gives rise to adverse degradation effects on mechanical properties of structural materials in reactors under irradiation. Multi-scale modeling scheme has been developed to study effects of helium on the long-term microstructural evolution. However, the implementation of Arrhenius law based on the quasi-equilibrium reaction process  is not appropriate to predict the migration behavior of helium in metals due to low-energy barrier. A coarse-grained formula is required to incorporate the non-equilibrium nature, e.g., the dissipative friction coefficient $\gamma$. In this paper, we derive an analytical expression for $\gamma$ based on a coarse-grained model of Brownian motion upon a periodic potential, in terms of dissipative feature of the thermal excitations in the many-body dynamical system by constructing an adiabatic relaxation process, which is then confirmed by a numerical example of vacancy migration in BCC W. Then the many-body dynamics simulations are performed for helium migration in BCC W and Fe, where the classical mobility are obtained and in good agreement with the data from experiments and other calculations. Finally, we propose a coarse-grained formula for the helium migration in BCC W and Fe, i.e., Eq.~(\ref{Eq.:coarse-grained}) in the context, using the calculated parameters from the adiabatic relaxation simulations. This work would help to develop a new multi-scale modeling scheme for effects of helium in metals, as well as the atomistic reactions with low-energy pathways in materials science.   
\end{abstract}


\end{frontmatter}

\section{\label{Sec.1}{Introduction}}

	Helium (He) is one of the most common productions in structural materials of fission and fusion reactors under irradiation. Due to the insolubility of He atom in metals, it would be easily trapped into sinks, such as vacancies and grain-boundary, and would form the helium-bubble in a long-term evolution process. The formation and accumulation of helium lead to the adverse ageing effects on structural materials, typically the high-temperature helium embrittlement \cite{duffy2011modelling,zinkle2013materials}. Therefore, understanding the kinetics of the bubble nucleation and growth is considered as one of the key issues in nuclear materials science and engineering \cite{TRINKAUS2003229}. 
	
\begin{figure}
			\graphicspath{}
			\makeatletter
			\def\@captype{figure}
			\makeatother
			\centering
			\includegraphics[width=0.45\textwidth]{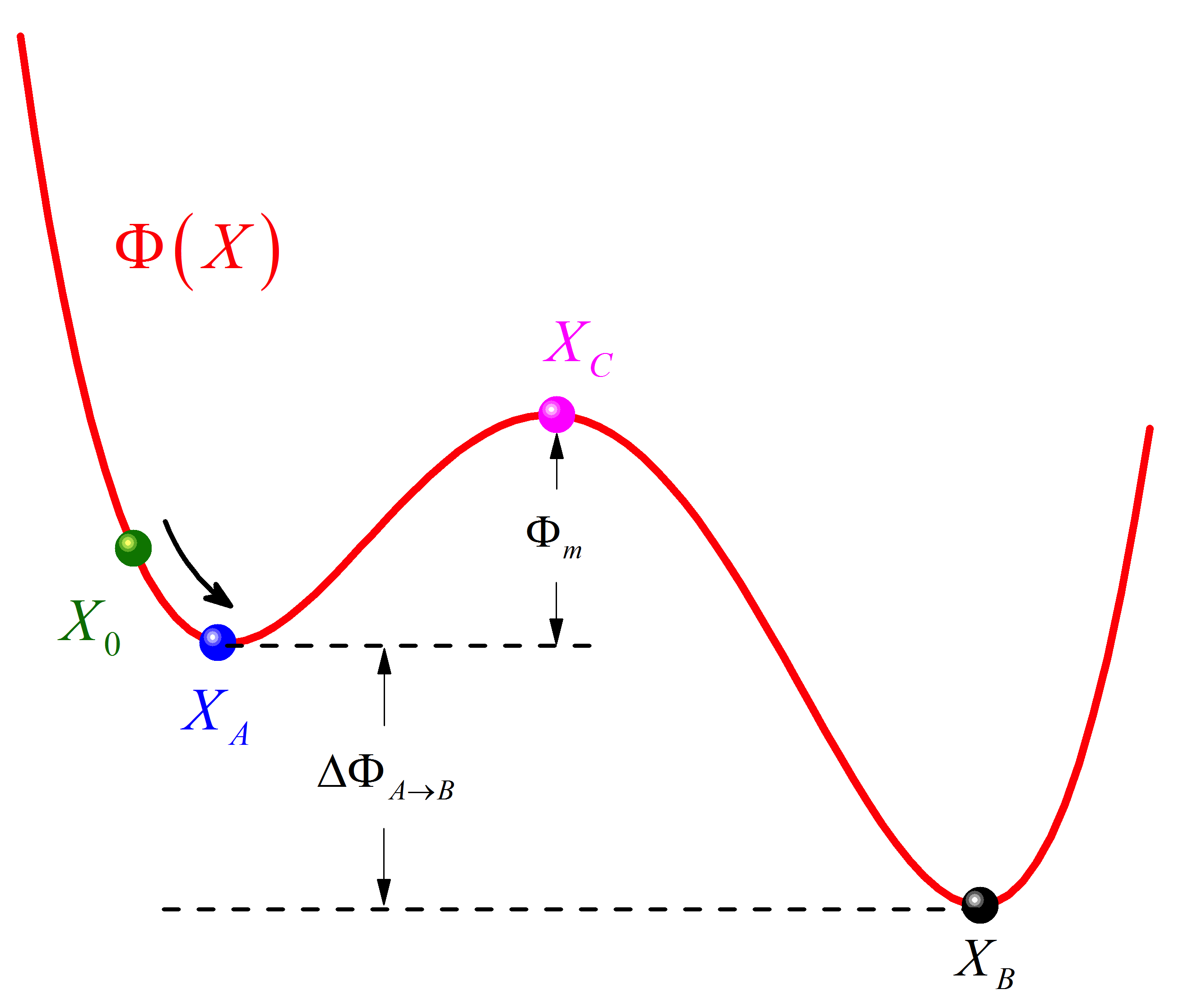}
			\caption{A universal mesoscale coarse-grained model of atomic activation, with the horizontal and vertical axes respectively the reaction coordinate $X$ and energy $\Phi(X)$. The system moves from one stable state $X_A$ to another $X_B$, by going over the saddle-point state $X_C$, where $\Phi_m$ is the activation energy, and $\Delta \Phi_{A\rightarrow B}$ is the energy difference between states of $X_A$ and $X_B$. $X_0$ denotes an excited state near $X_A$. }
			\label{Fig.1}
		\end{figure}

	As a long-term phenomenon, the formation and accumulation of helium consist of a large numbers of atomic activation processes, covering from electronic scale to macroscale in spatial-scale and femto-second to decades in temporal-scale. Multi-scale modeling scheme has been well-established to study the microstructural evolution arising from effects of helium formation and accumulation \cite{samaras_multiscale_2009}. In this scheme, each individual activation is regarded as an atomistic reaction as schematic as in Fig.~\ref{Fig.1}, with the activation rate $\nu$ determined using Arrhenius law, 
		\begin{equation}\label{Eq.:Arrhenius_1}
			\nu = \nu_0 e^{-G/k_\textrm{B}T} = \nu_0 e^{S/k_\textrm{B}} e^{-\Phi_m/k_\textrm{B}T}
		\end{equation}
	Here, $\nu_0$ is the attempt frequency, denoting the equilibration temporal characteristics; $G = \Phi_m - TS$ is the free energy barrier with $\Phi_m$ and $S$ respectively the energy and entropy; $k_\textrm{B}$ is Boltzmann constant and $T$ is absolute temperature. Note that Arrhenius law is only applicable for a quasi-equilibrium process, i.e., the reaction with $\mathcal{T}$$\ll$1, where $\mathcal{T} = k_\textrm{B}T/\Phi_m$ is the defined effective temperature. For reactions with low-energy pathway, it undergoes a non-equilibrium process when $\mathcal{T}$$\gg$1, which could not be described by Eq.~(\ref{Eq.:Arrhenius_1}) \cite{dudarev2008non}. In this case, a dissipative friction coefficient $\gamma$ is introduced to denote the temporal characteristics of a non-equilibrium reaction, which should be incorporated into the multi-scale modeling schemes \cite{henkelman_atomistic_2017}. 
	
	As a basic requirement of bubble nucleation and growth \cite{trocellier_review_2014}, interstitial helium diffusion in metals is a typical low-energy reaction. Calculations based on density functional theory (DFT) \cite{PhysRevLett.97.196402} found that $\Phi_m$$\sim$0.1eV of helium in metals. Molecular dynamics (MD) simulations \cite{perez2014diffusion,wen_many-body_2017} found that helium in BCC W reveals Arrhenius-type quasi-equilibrium diffusion at $T$$<$500K, i.e., Eq.~(\ref{Eq.:Arrhenius_1}), and Einstein-type non-equilibrium diffusion at $T$$>$700K, i.e., $D\propto T$. A stochastic model based on a simplified saw-tooth potential for Brownian motion indicates such non-Arrhenius diffusion behavior is arising from the competition between the stochastic force from phonon-wind and the restoring force from the conserved crystal potential \cite{wen_many-body_2017}. Note that the reaction behaviors with low-energy pathway in low- and high-temperature limits are well-understood, respectively, the key problem is how to appropriately describe the reaction behavior at $\mathcal{T}$$\approx$1, which has been well studied \cite{sancho2004diffusion, pavliotis2008diffusive} based on Kramers' theory \cite{KRAMERS1940284}, and applied to other related areas, such as surface-diffusion \cite{ALANISSILA1992227, lechner2013atomic,krylov2014physics} and dislocation motion in metals \cite{dudarev2002thermal,DUDAREV2002881,PhysRevB.84.134109,Swinburne2014}. In these studies, a coarse-grained formula of the classical mobility $\mu$ of low-energy reaction is derived to incorporate the equilibration $\nu_0$ and dissipation temporal characteristics $\gamma$ within a unique framework. However, relevant investigations on helium-induced microstructural evolution are still incomplete, in particular about the issue how to obtain $\gamma$ from the many-body stochastic environment, which is required beforehand for the implementation of existing models. 
	
	In this paper, aiming to the point-defect diffusion in metals, we derive an analytical expression of $\gamma$ in terms of dissipative features of thermal excitations of the stochastic many-body system by constructing an adiabatic relaxation process for an atomistic reaction, and get the microdynamic insight of the non-equilibrium nature of thermal-assisted reaction phenomena, where vacancy migration in BCC W is taken as an example to check the validity in Sec.~\ref{Sec.2}. Then, we apply this approach to the interstitial helium migration in BCC W and Fe to calculate the classical mobility. Using the parameters obtained in the adiabatic relaxation process as input, a coarse-grained formula is proposed based on a modified Brownian diffusion model upon a sinusoidal-type potential to describe the kinetics of interstitial diffusion of helium in mesoscale in Sec.~\ref{Sec.3}. The paper is concluded in Sec.~\ref{Sec.4}. 

\section{\label{Sec.2}Theoretical Model}

	\subsection{Adiabatic relaxation process for $\gamma$}\label{Sec.2.1}
	As schematic in Fig.~\ref{Fig.1}, the many-body dynamics of crystalline solid containing point defects is coarse-grained as the one-particle motion in mesoscale, i.e., one-dimensional Brownian motion upon a periodic potential, and governed by a generalized Langevin equation (GLE) 
		\begin{equation} \label{Eq.:Langevin Equation}
			m^* \ddot{X} = -\partial_X \Phi(X) - m^*\gamma \dot{X} + f(t) \ ,
		\end{equation}
	where $X$ is the mesoscale coordinate of the coarse-grained many-body system; $m^*$ and $\gamma$ are respectively its effective mass and friction coefficient, from which the thermal drag mobility is thus defined as $\mu_d$=$1/\left( m^*\gamma\right)$; $-\partial_X \Phi(X)$ is the restoring force along the reaction path provided by the periodic conserved crystal potential $\Phi(X)$; $f(t)$ is a Gaussian random force. Note that, all the terms in Eq.~(\ref{Eq.:Langevin Equation}) intrinsically have the many-body nature, which are coarse-grained quantities representing a large number of microscopic degrees of freedom in atomic system. 
	
	By multiplying $\dot{X}$ at both sides of Eq.~(\ref{Eq.:Langevin Equation}), we have 
		\begin{equation} \label{Eq.:Work_done}
			\begin{aligned}
				m^*\ddot{X}\dot{X} &= -\dot{X}\partial_X \Phi - m^*\gamma \dot{X}^2 + f(t)\dot{X} \\
				\Rightarrow \frac{\textrm{d}U(t)}{\textrm{d}t} & = -2\gamma K(t) + \eta(t) \\
			\end{aligned}
		\end{equation}
	where $U = \Phi + K$ is the total energy with $K(t)$$=$$m^*\dot{X}^2/2$ the kinetic energy, $\eta(t)$$=$$f(t)\dot{X}$ is the instantaneous work-done rate of the random force. Seen from Eq.~(\ref{Eq.:Work_done}), $\gamma$ could be directly calculated from the dissipation rate of total energy without attached to the thermal environment. As shown in Fig.~\ref{Fig.1}, an adiabatic relaxation process is constructed by initializing the system at an excited state $X_0$ near $X_A$ with zero velocity $\dot{X}(0)$$=$$0$ and de-attaching its thermal environment, i.e., $f(t)$=0 in Eq.~(\ref{Eq.:Langevin Equation}). In this case, the system relax towards $X_A$ adiabatically like a damped oscillator governed by 
		\begin{equation} \label{Eq.:Vac_EOM}
			m^* \ddot{X} + m^* \gamma \dot{X} + m^* \omega_A^2 X = 0
		\end{equation}
	where the conserved force field $\Phi(X)$ is assumed to be expanded harmonically near $X_A$ as
		\begin{equation}
			\begin{aligned}
				\Phi(X) - \Phi(X_A) & = \frac{1}{2}m^* \omega_A^2 (X - X_A)^2 \\
				\Rightarrow \quad U^0 = \Phi(X_0) & = \frac{1}{2} m^* \omega_A^2 X_0^2 \\
			\end{aligned}
		\end{equation}	
	where $\omega_A$ is the vibrational frequency and $U^0$ is the initial energy, and setting $X_A$$=$$0$ and $\Phi(X_A)$$=$$0$ for convenience. Neglecting the oscillating behavior, the energy $U(t)$ at the limit of $\gamma\ll\omega_A$ reduces following
		\begin{equation}\label{Eq.:Ut_macro}
			U(t) = U^0  e^{- \gamma t} = U^0  e^{-t/\tau}
		\end{equation}
	The characteristic dissipation time $\tau$ is thus given by
		\begin{equation}\label{Eq.:tau_1}
			\tau = \gamma^{-1} = \int_{0}^{\infty} \frac{U(t)}{U^0} \textrm{d}t
		\end{equation}
	Here, the validity of the approximation $\gamma\ll\omega_A$ in the deduction of Eq.~(\ref{Eq.:Ut_macro}) and (\ref{Eq.:tau_1}) is discussed in \ref{Sec.A1}. 	
	
	Following Zwanzig \cite{Zwanzig1960} and Mori \cite{Mori1965} projection operator approach, the meso-scale coordinate could be projected to the hyperspace of microdynamic system \cite{Swinburne2014}, with $3N$-dimensional vector $\bm{X}$ on the basis of a set of atomic positions 
		\begin{equation}
			\bm{X} = \left\lbrace \bm{u}_1 \otimes\bm{u}_2 \otimes \cdots \otimes \bm{u}_N \right\rbrace \equiv \left\lbrace \otimes \bm{u}_l \right\rbrace \in \mathbb{R}^{3N} 
		\end{equation}
	where $\bm{u}_l$ is the $l\textrm{th}$ atomic displacement with the unit-vector $\mathbf{e}_l$. On the other hand, since phonon-coordinates system in reciprocal space is an equivalent complete set to the corresponding atomic coordinates system in real space, $\bm{X}$ could be alternatively projected to the phonon-space as
		\begin{equation} \label{Eq.:projection_chain}
			\begin{aligned}
				\bm{X} =\left\lbrace \otimes \bm{u}_l \right\rbrace 
					= \left\lbrace \otimes \xi_{k\sigma} \right\rbrace 
					= \left\lbrace \otimes n_{k\sigma} \right\rbrace \in \mathbb{R}^{3N} 
			\end{aligned}
		\end{equation}
	where $\xi_{k\sigma}$ is the coordinate of phonon mode $\left( k,\sigma \right)$ with the unit-vector $\mathbf{e}_{k\sigma}$, and $n_{k\sigma}$ is its occupation number. In this regard, the movement of phase-point in phase-space corresponds to the evolution of a set of coordinates $\left\lbrace n_{k\sigma}\right\rbrace $ in hyperspace of phonon modes with basis as $\left\lbrace \hbar\omega_{k\sigma}\right\rbrace$.  
	
	Correspondingly, an adiabatic relaxation process mentioned above is actually the momentum and energy transfer process of phonon modes in a many-body system due to the intrinsic anharmonic effects, arising from either the scattering by crystalline defects or phonon-phonon collisions, which could be denoted as phonon creation or annihilation. Accordingly, the occupation number $\left\lbrace n_{k\sigma}(t) \right\rbrace$ varies with characteristic time $\tau_{k\sigma}$$=$$\left( 2\Gamma_{k\sigma}\right) ^{-1}$ ( $\Gamma_{k\sigma}$ is the spectral width)  \cite{VanKampen1992}
		\begin{equation} \label{Eq.:nk_t}
			\begin{aligned}
				n_{k\sigma}(t) = \left[ n_{k\sigma}^0 - n_{k\sigma}^\infty \right] e^{-t/\tau_{k\sigma}} + n_{k\sigma}^\infty 
			\equiv n_{k\sigma}^\infty \left( 1 - \delta_{k\sigma} e^{-t/\tau_{k\sigma}} \right) 
			\end{aligned}
		\end{equation}
	from initial state $\left\lbrace n_{k\sigma}^0 \right\rbrace $ at $t$$=$$0$ to its equilibrium state $\left\lbrace n_{k\sigma}^\infty \right\rbrace $ at $t$$\rightarrow$$\infty$, i.e., Bose-Einstein distribution,	
		\begin{equation}
			n_{k\sigma}^\infty = \left\langle n_{k\sigma}\right\rangle _T = 
				\left( e^{\hbar {\omega}_{k\sigma}/k_\textrm{B}T^\infty} - 1 \right)^{-1} \approx  \frac{k_\textrm{B}T^\infty}{\hbar {\omega}_{k\sigma}}
		\end{equation}	
	in the classical limit, with $T^\infty$ the equilibrium temperature of the fully relaxed phonon system. Here, $\delta_{k\sigma}$ represents the \emph{relative distance} of phonon mode $\left(k,\sigma\right)$ from initial state and equilibrium, 
		\begin{equation}\label{Eq.:delta_ks}
			\delta_{k\sigma} \equiv 1 - \left(n_{k\sigma}^0/ n_{k\sigma}^\infty\right) = 1 - \left(\varepsilon_{k\sigma}^0/ \varepsilon_{k\sigma}^\infty\right)
		\end{equation}
	with $\varepsilon_{k\sigma}$$=$$n_{k\sigma}\hbar {\omega}_{k\sigma}$ the phonon energy. Further, the total energy keeps constant during an adiabatic process, giving rise to $\sum_{k\sigma} \delta_{k\sigma} = 0$.	
	
	In addition, according to the definition of entropy in classical limit, i.e., $S_{k\sigma}$$=$$k_\textrm{B} \left( \ln n_{k\sigma} + 1 \right)$, the entropy production $\Delta S_{k\sigma}$ for phonon mode $\left(k,\sigma \right) $ relaxed towards equilibrium can be written as
		\begin{equation} \label{Eq.:s_ks}
			\Delta S_{k\sigma} \equiv S_{k\sigma}^\infty -  S_{k\sigma}^0 = - k_\textrm{B} \ln \left( 1 - \delta_{k\sigma}\right)
		\end{equation}	
	Here, $n_{k\sigma}^0 \geqslant 0$, so that $\left(1 - \delta_{k\sigma}\right) \geqslant 0$, satisfying the requirement of logarithm function in Eq.~(\ref{Eq.:s_ks}). Therefore, the total entropy production $\Delta S$ is 	
		\begin{equation} \label{Eq.:S_system}
			\begin{aligned}
				\Delta S & 
				= -\sum_{k\sigma} k_\textrm{B} \ln \left(1 - \delta_{k\sigma}\right) 
				= -k_\textrm{B} \ln \left\lbrace \prod_{k\sigma} \left( 1- \delta_{k\sigma} \right) \right\rbrace \\
				& \geqslant - k_\textrm{B} \ln \left\lbrace \left[ \frac{1}{N}\sum_{k\sigma} \left( 1- \delta_{k\sigma} \right)\right] ^N \right\rbrace = 0 \\
			\end{aligned}			
		\end{equation}	
	where the Cauchy inequality is applied. Here, $\Delta S$$\geqslant$$0$ is consistent with the 2nd-law of thermodynamics, which indicates the expression of Eq.~(\ref{Eq.:nk_t}) is appropriate to describe the microdynamics during an adiabatic process of phase transport with maximizing entropy. The equal sign in Eq.~(\ref{Eq.:S_system}) holds with  $\forall\delta_{k\sigma}$$=$$0$, giving rise to $n_{k\sigma}^0$$=$$n_{k\sigma}^\infty$, which means the initial state is the equilibrium state. In other word, $\delta_{k\sigma}$$\neq$$0$ is the intrinsic driving force of the heat dissipation by the re-distribution of phonon modes in the many-body system. 
	
	Note that, $U(t)$ in Eq.~(\ref{Eq.:Ut_macro}) is indeed the free energy $A(t)$ of the corresponding phonon system, which thus converts into heat $Q(t)$ since the internal energy $E$ keeps constant during the adiabatic process, so that 	
		\begin{equation}
			\dot{U}(t) = \dot{A}(t) \equiv \dot{E}(t) - \dot{Q}(t) = -\dot{Q}(t)
		\end{equation}		
	Then, we have
		\begin{equation} \label{Eq.:U_and_Q}
			\begin{aligned}
				e^{-\gamma t} = \frac{U(t) - U^\infty}{U^0 - U^\infty} = \frac{Q^\infty - Q(t)}{ 3Nk_\textrm{B}T^\infty} = \frac{\int_{t}^{\infty} \dot{Q}(t') \textrm{d} t'}{3Nk_\textrm{B}T^\infty} 
			\end{aligned}
		\end{equation}	
	where $U^0$$=$$Q^\infty$$\equiv$$3Nk_\textrm{B}T^\infty$ and $U^\infty$$=$$Q^0$$=$$0$. In addition, the heat dissipation rate $\dot{Q}(t)$ corresponds to the rate of phonon mode re-distribution, equivalent to the \emph{velocity} of microdynamic coordinate in phonon-space	
		\begin{equation}\label{Eq.:Q_t}
			\begin{aligned}
				\dot{Q}(t) & = \sum_{k\sigma} T_{k\sigma}  \dot{S}_{k\sigma} 
				= \sum_{k\sigma} k_\textrm{B}T_{k\sigma} \frac{\textrm{d}}{\textrm{d} t} \left[   \ln n_{k\sigma}(t) \right] \\
				& = \sum_{k\sigma} \dot{n}_{k\sigma}(t)\hbar {\omega}_{k\sigma}  
				= \sum_{k\sigma} \dot{\varepsilon}_{k\sigma}(t)  \\
			\end{aligned}			
		\end{equation}	
	where $k_\textrm{B}T_{k\sigma}$$\equiv$$\varepsilon_{k\sigma}$ is the defined local temperature in phonon space. Substituting Eq.~(\ref{Eq.:Q_t}) into Eq.~(\ref{Eq.:U_and_Q}), 		
		\begin{equation}
			\begin{aligned}
				e^{-\gamma t} 
					& = \frac{\int_{t}^{\infty} \dot{Q}(t') \textrm{d} t'}{3Nk_\textrm{B}T^\infty} 
					= \frac{1}{3N}\sum_{k\sigma} \left[ \frac{\varepsilon_{k\sigma}^\infty - \varepsilon_{k\sigma}(t) }{\varepsilon_{k\sigma}^\infty} \right] \\
					& = \frac{1}{3N} \sum_{k\sigma}  \delta_{k\sigma}  e^{-t/\tau_{k\sigma}} \\
			\end{aligned}
		\end{equation}	
	The characteristic dissipation time $\tau$ is thus obtained following Eq.~(\ref{Eq.:tau_1}) as
		\begin{equation} \label{Eq.:Mid_gG}
			\tau = \gamma^{-1}  = \frac{1}{3N} \sum_{k\sigma} \delta_{k\sigma}\tau_{k\sigma} 
				= \left\langle \delta_{k\sigma}\tau_{k\sigma} \right\rangle
		\end{equation}	
	Note that the kinetic energy of phonon system shares the same dissipative temporal characteristics as heat dissipation, because
		\begin{equation}\label{Eq.:K_t}
			\dot{K}(t) = \left\langle k_\textrm{B}\dot{T}_{k\sigma}\right\rangle = \left\langle  \dot{n}_{k\sigma}(t) \hbar\omega_{k\sigma}\right\rangle \propto \dot{Q}(t)
		\end{equation}
	In this regard, the dissipative rate of $K(t)$ could be used to estimate the friction coefficient $\gamma$ in Eq.~(\ref{Eq.:Langevin Equation}). 
	
	Here, we derive an analytical expression of the coarse-grained dissipative friction coefficient of point defect diffusion in mesoscale, in terms of the characteristic relaxation time $\left\lbrace \tau_{k\sigma}\right\rbrace$ and the \emph{relative distance} $\left\lbrace \delta_{k\sigma}\right\rbrace$ away from equilibrium of the phonon modes in a microscale, bridging the mesoscale kinetics and microscale many-body dynamics.

	\subsection{Numerical example: vacancy diffusion in W}

\begin{figure}
		\graphicspath{../Figures}
		\makeatletter
		\def\@captype{figure}
		\makeatother
		\centering
		\includegraphics[width=0.48\textwidth]{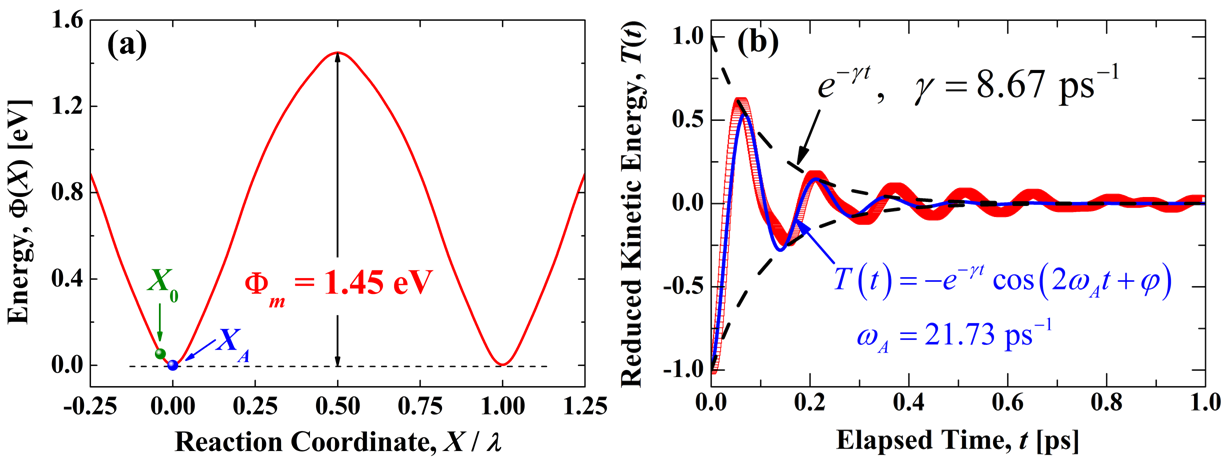}
		\caption{\label{Fig.2} (a) The energy profile $\Phi(X)$ of migration process for vacancy in BCC W along the reaction path with $\lambda$=2.47\AA, calculated using modified conjugated gradient method \cite{Sinclair1974}. Here, $\Phi_m$=1.45eV is the migration energy, $X_A$ represents one equilibrium state, and $X_0$ is an excited state near $X_A$ along the diffusion path. (b) The reduced kinetic energy $T(t)$ of the corresponding adiabatic relaxation process. Here, the red rectangle points are the simulation results, fitting by $T(t) = -e^{-\gamma t}\cos(2\omega_A t + \varphi)$ (blue line), with $\gamma$=8.67/ps and $\omega_A$=21.73/ps, and black dashed line is the envelop of $\pm e^{-\gamma t}$.}
	\end{figure}

	In the following, we will examine the above expression Eq.~(\ref{Eq.:Mid_gG}) by taking vacancy migration in BCC W as an example. As a thermodynamically stable point-defect at finite temperatures, a vacancy can jump from an equilibrium state to a neighboring one by going over an energy barrier with the help of phonon-scattering. Plotted in Fig.~\ref{Fig.2}(a), the migration energy $\Phi_m$ along the reaction path is $\sim$$1.45$eV, so that  $\Phi_m$$\gg$$k_\textrm{B}T$ when $T$$<$$T_m$ with the melting point $T_m$$\sim$3800K for BCC W. Therefore, vacancy migration is a typical Brownian motion upon a periodic potential, which can be described by GLE in Eq.~(\ref{Eq.:Langevin Equation}). 
	
	In the many-body atomistic model, vacancy is mimicked by an ensemble of $N$ atoms located in ($N$+1)  lattice sites, where the simulation box  includes $10$$\times$$10$$\times$$9$ BCC unit-cells in Cartesian coordinate system with periodic boundary condition applied to avoid the surface effects, and Ackland's potential \cite{Ackland:1987} adopted to describe the W-W interatomic interaction. The many-doby Hamiltonian $\mathscr{H}$ is written as 
		\begin{equation}
			\mathscr{H} = \mathscr{H}_0 + \mathscr{H}'  
				= \sum_{l}\frac{\bm{p}_l^2}{2m_l} + \Phi_0\left(\left\lbrace \bm{R}_l \right\rbrace\right)  + \mathscr{H}'
		\end{equation}
	where $\bm{p}_l$, $\bm{R}_l$ and $m_l$ are respectively the atomic momentum, position and mass of the $l\textrm{th}$ atom; $\mathscr{H}_0$ corresponds to the Hamiltonian of a perfect crystal, with $\Phi_0$ the interatomic potential with respect to the atomic configuration $\left\lbrace \bm{R}_l \right\rbrace $, and $\mathscr{H}'$ represents the effect of vacancy. In molecular dynamics simulations, $\mathscr{H}'$ is incorporated into the interatomic interaction with a missing atom inside, which could be seen clearly in the derived equations of motion
		\begin{equation} \label{Eq.:atom_EOM}
			\begin{aligned}
				\frac{\textrm{d}\bm{R}_l}{\textrm{d}t} = \frac{\bm{p}_l}{m}, \quad \textrm{and} \quad
				\frac{\textrm{d}\bm{p}_l}{\textrm{d}t} = -\frac{\partial \Phi'}{\partial \bm{R}_l}			
			\end{aligned}
		\end{equation}
	with $\Phi'=\Phi_0+\mathscr{H}'$. 
	
	Following the procedure mentioned in Sec.~\ref{Sec.2.1}, the many-body system adiabatically relaxes starting from the initial state with zero-temperature, i.e., $\bm{p}_l$=0, and the microdynamic phase-space trajectory $\left\lbrace \bm{p}_l, \bm{R}_l\right\rbrace$ is recorded by solving the equation of motion in Eq.~(\ref{Eq.:atom_EOM}). The kinetic energy $K(t)$ is then calculated following   		
		\begin{equation} \label{Eq.:Kt}
			K(t) = \sum_{l} \frac{1}{2} m \dot{\bm{R}}_l^2(t) = \frac{1}{2} \sum_{k\sigma} \varepsilon_{k\sigma}(t)
		\end{equation} 	
	Note that, the potential energy converts into the kinetic energy in an adiabatic process, which are balanced at equilibrium state, so that $K(t)\arrowvert_{t\rightarrow\infty}$$=$$K^\infty$$=$$U_0/2$. According to Eq.~(\ref{Eq.:K_t}), the reduced kinetic energy $T(t)$ is defined to obtain the relaxation information. Plotted in Fig.~\ref{Fig.2}(b), $T(t)$ behaves like an underdamped oscillator, which could be described as
		\begin{equation} \label{Eq.:Ek_damping}
			T(t) = \left( \frac{K(t)}{K^\infty} - 1 \right) = - e^{-\gamma t} \cos\left( 2\omega_A t + \varphi\right) 
		\end{equation} 	
	By fitting the simulation data of $T(t)$ in Fig.~\ref{Fig.4} following Eq.~(\ref{Eq.:Ek_damping}), we have		
		\begin{equation}\label{Eq.:gamma_relax}
			\gamma = 8.67/\textrm{ps} , \quad  \omega_A = 21.73/\textrm{ps} \quad \textrm{and} \quad  \varphi = 0
		\end{equation}	
	giving rise to $\tau = \gamma^{-1} \approx 0.115 \textrm{ps}$. 
	
\begin{figure}[tpb]
		\graphicspath{}
		\makeatletter
		\def\@captype{figure}
		\makeatother
		\centering
		\includegraphics[width=0.4\textwidth]{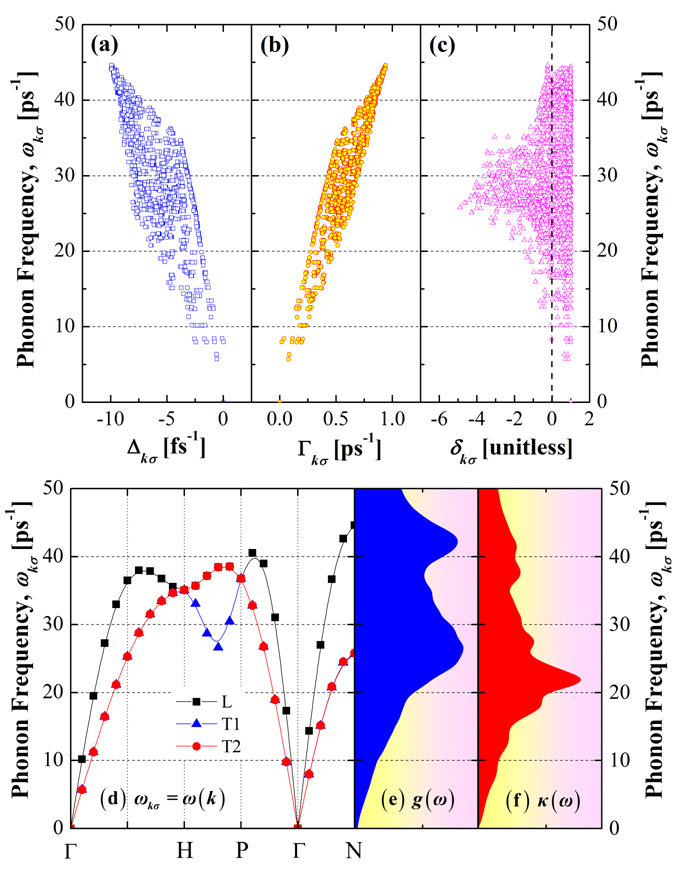}
		\caption{The (a) frequency-shift $\Delta_{k\sigma}$ and (b) spectral width $\Gamma_{k\sigma}$ of the characteristic phonon modes $\left(k,\sigma\right)$ with the impact of vacancy in BCC W, and (c) the \emph{relative distance} $\delta_{k\sigma}$ of phonon mode along the reaction path in a constructed adiabatic relaxation process of phase transport; (d) The phonon dispersion relation $\omega_{k\sigma} = \omega(k)$ and (e) density of states of perfect BCC W and (f) the phonon modes spectrum $\kappa(\omega)$ participated in the phase transport, obtained by applying Fourier transform on $T(t)$ in Fig.~\ref{Fig.2}(b).  Here $L$ (black squares), $T_1$ (blue triangles) and $T_2$ (red solid-circles) represent the three branches of phonon mode for a given wave-vector.}
		\label{Fig.3}
	\end{figure}

\begin{figure}
		\graphicspath{}
		\makeatletter
		\def\@captype{figure}
		\makeatother
		\centering
		\includegraphics[width=0.48\textwidth]{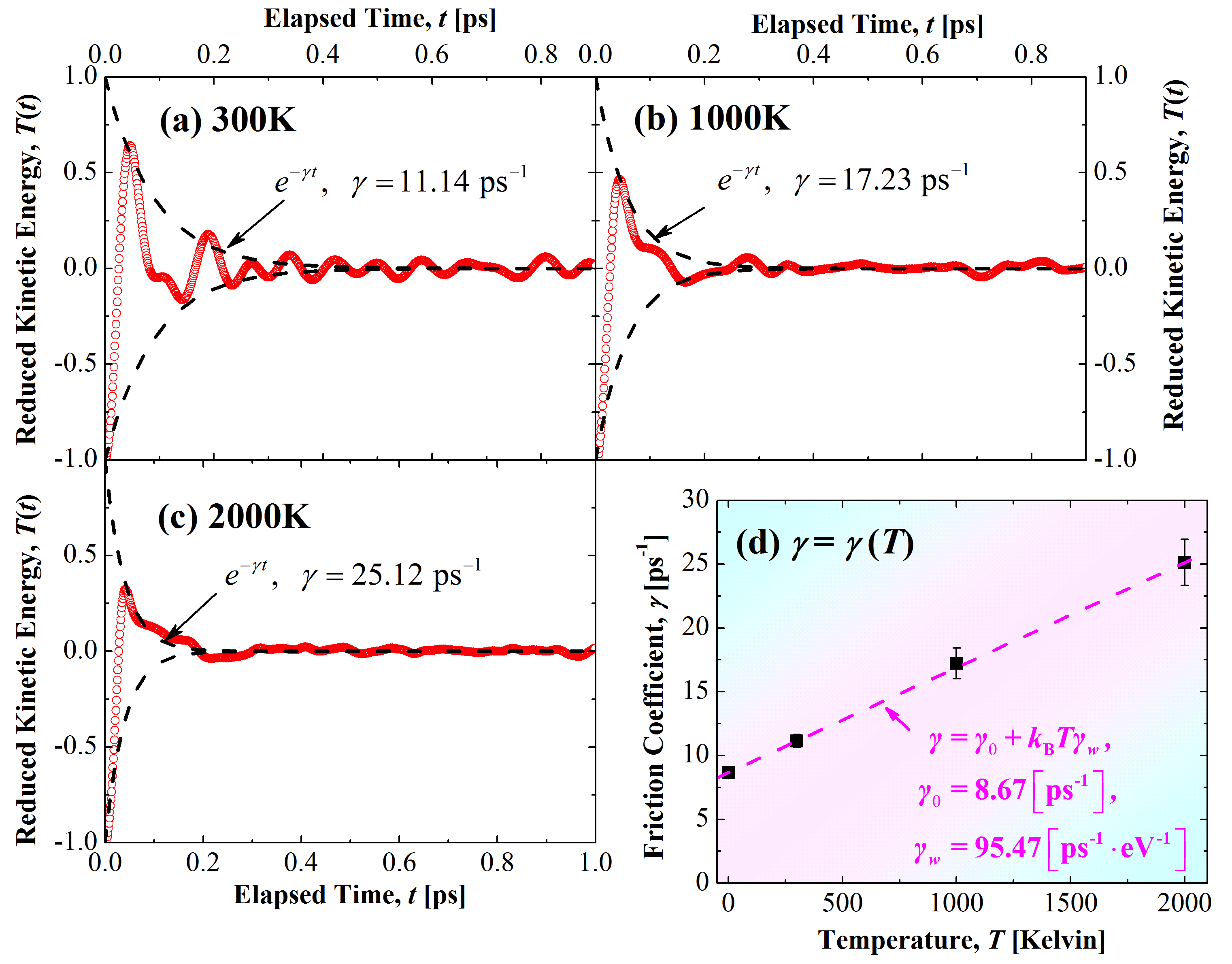}
		\caption{The reduced kinetic energy $T(t)$ of an adiabatic relaxation process for vacancy migration in BCC W at various initial background temperatures, i.e., (a) 300K, (b) 1000K and (c) 2000K, and (d) the corresponding friction coefficient $\gamma$ is plotted as an linear function of temperature, indicating similar behavior discussed in Ref.~\cite{Swinburne2014}, where $\gamma = \gamma_0 + k_\textrm{B}T\gamma_w$ with $\gamma_0$=8.67/ps and $\gamma_w$=95.47/$\left( \textrm{ps}\cdot \textrm{eV}\right)$.}\label{Fig.4}
	\end{figure}

	This many-body dynamical system could be also treated as an ensemble of phonon modes perturbed by vacancy, with Hamiltonian $\mathscr{H}$ as 	
		\begin{equation} \label{Eq.:2.1}
			\mathscr{H} = E_0 + \sum_{k\sigma} {n_{k\sigma} \hbar \omega_{k\sigma}}  + \mathscr{H}' \equiv E_0 + \sum_{k\sigma} {n_{k\sigma} \hbar \tilde{\omega}_{k\sigma}}
		\end{equation}	
	where $E_0$ is the static energy with all the atoms at their equilibrium positions; $\omega_{k\sigma}$ and $\tilde{\omega}_{k\sigma}$ are the phonon frequencies without and with vacancy perturbation, respectively; $n_{k\sigma}$$=$$a_{k\sigma}^+ a_{k\sigma}$ is the occupation number, with $a_{k\sigma}^+$ and $a_{k\sigma}$ respectively the creation and annihilation operators. Here, effect of $\mathscr{H}'$ is involved in $\tilde{\omega}_{k\sigma}$, giving rise to frequency-shift $\Delta_{k\sigma}$ and spectral-width $\Gamma_{k\sigma}$, as $\tilde{\omega}_{k\sigma} = \omega_{k\sigma} + \Delta_{k\sigma} - \textrm{i}\Gamma_{k\sigma}$ (See in \ref{Sec.A2}). In atomic simulation, $\omega_{k\sigma}$ and $\tilde{\omega}_{k\sigma}$ could be respectively obtained by solving the following eign-equations, 
		\begin{equation}			
				\begin{aligned}
					\sum_{l\alpha} V_{l\alpha}^{n\beta}(k)\mathbf{e}_{l\alpha}  = \omega_{k\sigma}^2 \mathbf{e}_{n\beta}, \quad 
					\sum_{l\alpha} W_{l\alpha}^{n\beta}(k)\mathbf{e}_{l\alpha}  = \tilde{\omega}_{k\sigma}^2 \mathbf{e}_{n\beta} 
				\end{aligned}			
		\end{equation}	
	where $\mathbf{e}_{l\alpha}$ is the eign-vector of the $l\textrm{th}$ atom and $\alpha$ component; $V_{l\alpha}^{n\beta}(k)$ and $W_{l\alpha}^{n\beta}(k)$ the dynamic matrices of the systems without and with vacancy, respectively. Therefore, the frequency shift $\Delta_{k\sigma}$ is obtained by $ \Delta_{k\sigma}= \tilde{\omega}_{k\sigma} - \omega_{k\sigma}$, which are ploted in Fig.~\ref{Fig.3}(a). $\Delta_{k\sigma}$ is in the order of $\textrm{fs}^{-1}$, which is very tiny compared to $\tilde{\omega}_{k\sigma}$ (in the order of $\textrm{ps}^{-1}$). Accordingly, if $\left|\Delta_{k\sigma}\right|$$\ll$$\omega_{k\sigma}$, $\Gamma_{k\sigma}$ could be estimated by \cite{Fultz2010}
		\begin{equation}
			\Gamma_{k\sigma} \approx \sqrt{2 \tilde{\omega}_{k\sigma} \left| \Delta_{k\sigma}\right| }
		\end{equation} 
	Fig.~\ref{Fig.3}(b) shows $\Gamma_{k\sigma}$ is almost a linear function of $\omega_{k\sigma}$ with the slope $\sim$$0.02$, which indicates that the lattice distortion due to the vacancy results in a slight frequency-shift and spectral width of phonon modes.
	 
	In addition, the \emph{relative distance} $\left\lbrace \delta_{k\sigma}\right\rbrace $ for phonon mode relaxation during phase transport is determined by the initial configuration. Here, modified conjugated gradient (MCG) method \cite{Sinclair1974} is used to get the atomic configuration $\left\lbrace \bm{R}_l \right\rbrace $ of the many-body system with vacancy at equilibrium $\left\lbrace \bm{R}_l  \arrowvert{X_A}\right\rbrace$ and saddle-point states $\left\lbrace \bm{R}_l \arrowvert{X_C} \right\rbrace$ along the migratory direction, e.g., $\left\langle111\right\rangle$ in BCC crystal, with	$\Delta \bm{R}_{l}$$=$$\bm{R}_l \arrowvert{X_C}$$-$$\bm{R}_l \arrowvert{X_A}$ the $l\textrm{th}$ atomic displacement. The system is then initialized by setting $\left\lbrace \bm{R}_l \arrowvert{X_0} \right\rbrace$ as a small displacement $\left\lbrace \bm{u}_l\right\rbrace $ apart from $\left\lbrace \bm{R}_l  \arrowvert{X_A}\right\rbrace$ along the direction of $\left\lbrace\Delta \bm{R}_{l}\right\rbrace$, as	
		\begin{equation}
			\bm{R}_l \arrowvert{X_0} = \bm{R}_l  \arrowvert{X_A} + \bm{u}_{l} = \bm{R}_l  \arrowvert{X_A} + c \Delta \bm{R}_{l}
		\end{equation}	
	where $c$$=$$0.05$ is set in our simulation ($c$=0.1 is checked to show almost the same relaxation behavior). So that the initial projected phonon-coordinate $\xi_{k\sigma}^0$ is given by 
		\begin{equation}
			\xi_{k\sigma}^0 = \sum_{l} \sqrt{\frac{\hbar\tilde{\omega}_{k\sigma}}{Nm_l\tilde{\omega}_{k\sigma}^2}}\left( \bm{e}_{k\sigma} \cdot \bm{u}_l \right)  e^{-\textrm{i}\bm{k}\cdot \bm{R}_l}
		\end{equation}
	and the phonon energy $\varepsilon_{k\sigma}$ in harmonic approximation, 
		\begin{equation}
			\varepsilon_{k\sigma}^0 = n_{k\sigma}^0 \hbar \tilde{\omega}_{k\sigma} \approx \tilde{\omega}_{k\sigma}^2 \left| \xi_{k\sigma}^0\right|^2
		\end{equation}
	The \emph{relative distance} $\delta_{k\sigma}$ is then obtained by Eq.~(\ref{Eq.:delta_ks}). As plotted in Fig.~\ref{Fig.3}(c), phonon modes with $\omega_{k\sigma}$ ranging from 20/ps to 35/ps have large absolute values of $\delta_{k\sigma}$, while most of phonon modes has not been excited with $n_{k\sigma}^0$=0 leading to $\delta_{k\sigma}$=1, which are the principal modes participated in the relaxation process. Using the microdynamic information of $\tau_{k\sigma}$$=$$\left( 2\Gamma_{k\sigma}\right)^{-1}$ and $\delta_{k\sigma}$, the friction coefficient $\gamma$ is calculated following Eq.~(\ref{Eq.:Mid_gG}), 
		\begin{equation}\label{Eq.:gamma_micro_predict}
			\tau = \gamma^{-1} = \left\langle \delta_{k\sigma} \tau_{k\sigma} \right\rangle  \approx 0.102\ \textrm{ps} 
		\end{equation}
	which is in good agreement with the prediction from the calculated reduced kinetic energy in Fig.~\ref{Fig.2}(b) or Eq.~(\ref{Eq.:gamma_relax}). 
	
	Further, the oscillation behavior of $T(t)$ in Fig.~\ref{Fig.2}(b) corresponds to the thermal fluctuation of the many-body system near equilibrium. The oscillation frequency corresponds to the attempt frequency $\nu_0$ revealed in Arrhenius law, i.e., Eq.~(\ref{Eq.:Arrhenius_1}), as $\nu_0$$=$${\omega_A}/{2\pi}$=3.46/ps, which is consistent with the estimation using Vineyard's approach \cite{Vineyard1957}, as 
		\begin{equation} \label{Eq.:Vineyard}
			\nu_0 = \frac{1}{2\pi}\left( \prod\limits_{k}^{3N} \omega_k\right) \times \left( \prod\limits_{k}^{3N-1}\omega'_k\right) ^{-1} = 3.61 \ \textrm{ps}^{-1}
		\end{equation}	
	where $\omega_k$ and $\omega'_k$ are respectively the unconstrained and constrained eign-frequencies of vacancy system at equilibrium state $X_A$, calculated using modified conjugated gradient method \cite{Sinclair1974}. Here the constraint is applied to forbid the vibration along the reaction path. It is not surprised, because $\omega_A$ corresponds to the effective vibrational frequency of the vacancy system along its diffusion path as tackled in relaxation process, which is the inverse view with respect to Vineyard's approach in Eq.~(\ref{Eq.:Vineyard}). In addition, the effective mass $m^*$ of vacancy in a single one-dimensional jump process could be estimated in term of the curvature of energy profile $\Phi(X)$ and equilibration frequency $\omega_A$, as	
		\begin{equation}\label{Eq.:effective_mass}
			m^* = {\Phi''(X_A)} \times {\omega_A^{-2}} = 191.10 \ \textrm{a.u.}
		\end{equation}	
	which is comparable with the atomic mass of tungsten $m = 183.8 \ \textrm{a.u.}$. In this regard, the parameters required to set up a mesoscale generalized Langevin equation for the coarse-grained atomistic reaction, i.e.,  Eq.~(\ref{Eq.:Langevin Equation}), are directly obtained from an adiabatic relaxation process. 
	
	To sum up, the analytical expression of Eq.~(\ref{Eq.:Mid_gG}) is well-confirmed by the typical atomic activation of vacancy migration in BCC W, that the characteristic dissipation time $\tau$ of reaction corresponds to the life-time $\left\lbrace \tau_{k\sigma}\right\rbrace$ of phonon modes, weighted by the \emph{relative distance} $\left\lbrace \delta_{k\sigma}\right\rbrace$ along the reaction path. Compared to Dudarev's formulation \cite{DUDAREV2008,PhysRevB.92.134302}, there is no explicit term related to the effective mass in Eq.~(\ref{Eq.:Mid_gG}), because $\left\lbrace\Gamma_{k\sigma}\right\rbrace$ and $\left\lbrace\delta_{k\sigma}\right\rbrace$ do already include the characteristic features of the specific atomistic reaction. Let's look deep insight of the many-body dynamics during relaxation. Plotted in Fig.~\ref{Fig.3}(d) and (e), the phonon dispersion relation $\omega_{k\sigma}$=$\omega(k)$ and density of states $g(\omega)$ shows all the allowed phonon modes in BCC W. However, the spectral distribution of phonon modes excited during relaxation $\kappa(\omega)$ in Fig.~\ref{Fig.3}(f), the Fourier transform of $T(t)$ in Fig.~\ref{Fig.2}(b), just includes some of the allowed phonon modes, indicating that not all the  modes participate in the relaxation process. In this point of view, the presence of the  reaction-path-related $\left\lbrace \delta_{k\sigma}\right\rbrace$ in Eq.~(\ref{Eq.:Mid_gG}) plays an important role by providing the regulation for each phonon mode contributing to the mesoscale thermal-drag during the atomistic reaction process.
	
	In fact, $\gamma$ estimated from the adiabatic process with background temperature as zero corresponds to the athermal term $\gamma_0$ in Ref.~\cite{Swinburne2014}: ``arises because the defect displacement vector is not an eigenvector of the Hessian", which is consistent with our treatment by projecting ``the defect displacement" $X(t)$ on phonon-space with \emph{relative distance} $\left\lbrace \delta_{k\sigma}\right\rbrace$. In order to check this idea, relaxation processes are performed by initializing the system at various background temperatures $T^0$, i.e., upon adding an extra the atomic configuration of $\left\lbrace \left(  \bm{p}_l, \delta \bm{R}_l\right)\arrowvert T^0\right\rbrace$ obeying the corresponding Boltzmann distribution indexed by $T^0$ upon the initial configuration $\left\lbrace \bm{R}_l\arrowvert X_0 \right\rbrace$, giving rise to $X_0(T^0) = \left\lbrace \left( \bm{p}_l, \bm{R}_l+\delta \bm{R}_l \right) \arrowvert X_0, T^0\right\rbrace$. The reduced kinetic energy $T(t)$ with $T^0 =300, 1000, \textrm{and} \ 2000\textrm{K}$ are shown in Fig~\ref{Fig.4}, as well as the corresponding friction coefficient $\gamma$ as function of temperature. Here, $\gamma$ reveals a similar linear dependence on thermal energy, i.e., $\gamma = \gamma_0 + \gamma_w k_\textrm{B}T$ with $\gamma_w$=95.47/($\textrm{ps}$$\cdot$$\textrm{eV}$), which is believed to be arising from the high-order effects of phonon-wind.  

\section{\label{Sec.3}Classical mobility of He in BCC W and Fe}

	A single He atom occupying in the tetrahedral interstitial site of BCC metals induces a local strain field, and results in local resonance modes, as well as a scattering center of phonon modes. With the thermal fluctuations and the interatomic interactions provided by the host atoms, He atom exerts a periodic crystalline  potential and travels inside the metal until it is trapped by sinks. As shown in Fig.~\ref{Fig.5}(a), the migration energy $\Phi_m$ of interstitial helium in BCC W and Fe are very small, i.e., $\sim$0.145 and $\sim$0.091 eV, respectively, which are consistent with experimental data and other calculation results as listed in Table~\ref{tab:performance_comparison}. Therefore, helium migration is a low-energy atomistic reaction, which could be treated as the Brownian motion governed by GLE as in Eq.~(\ref{Eq.:Langevin Equation}). In this section, the classical mobility is firstly calculated from dynamical simulations as reference in Sec.~\ref{Sec.3.1}. Then simulations of the adiabatic relaxation process are performed to get the equilibration and dissipative features of helium diffusion in BCC W and Fe in Sec.~\ref{Sec.3.2}. Finally, a practical coarse-grained formula is proposed in Sec.~\ref{Sec.3.3} using the parameters obtained as input. 
	 
	\subsection{\label{Sec.3.1}Simulations of many-body dynamics}
		\begin{figure}
			\graphicspath{}
			\makeatletter
			\def\@captype{figure}
			\makeatother
			\centering
			\includegraphics[width=0.48\textwidth]{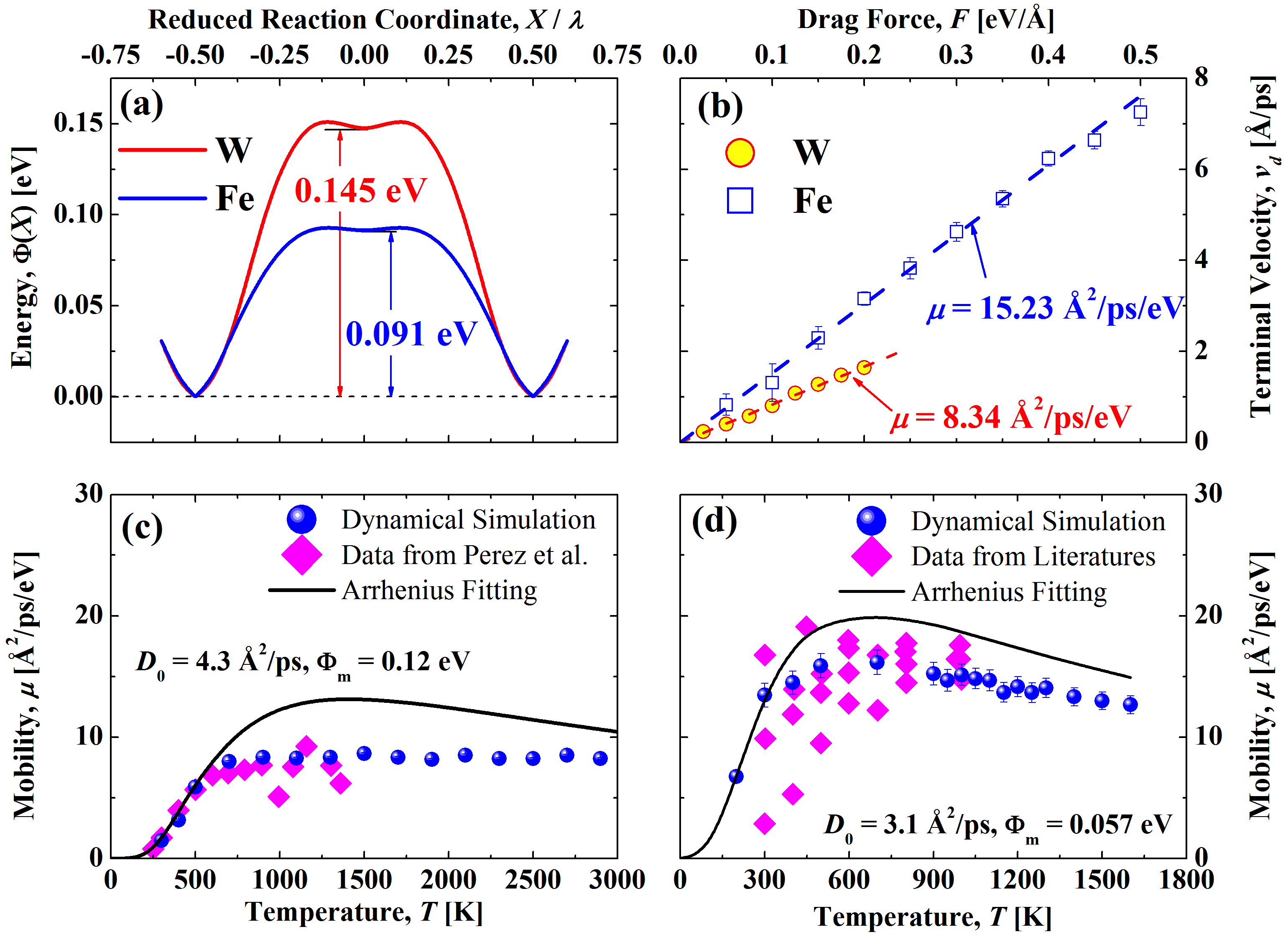}
			\caption{(a) The migratory energy profile of helium in BCC W and Fe; (b) demonstration of simulations of drift-dynamics on helium in BCC W and Fe at 900K, where the classical mobility is estimated. The calculated classical mobility of helium in BCC (c) W and (d) Fe, respectively, where the Arrhenius-fitting is also plotted with parameters shown inside. Here, the data from literature is illustrated for comparison, i.e., Perez et al. \cite{perez2014diffusion} in W, and Stewart et al. \cite{stewart_atomistic_2011} and references herein.}
			\label{Fig.5}
		\end{figure}
	The Hamiltonian $\mathscr{H}$ of a many-body system of a crystalline solid including an interstitial helium is written as
		\begin{equation}
			\mathscr{H} = \sum_{i=1}^{N} \frac{\bm{p}_l^2}{2m_l} + \Phi\left(\left\lbrace\bm{R}_l \right\rbrace  \right) 
		\end{equation}
	where $m_l$, $\bm{p}_l$ and $\bm{R}_l$ are respectively the $l$th atomic mass, momentum and coordinate; $\Phi\left(\left\lbrace\bm{R}_l \right\rbrace  \right)$ is the many-body interatomic potential. In this paper, potentials based on embedded atomic method (EAM)  are adopted to describe the interatomic interaction between W-W \cite{Ackland:1987,Juslin2013} , W-He \cite{Juslin2013}, Fe-Fe \cite{WEN2016102,Chiesa2011}, and Fe-He \cite{GAO2011115} atoms, respectively. The simulation box includes 20$\times$20$\times$20 BCC unit-cells with periodic boundary condition applied to avoid the surface effect. Using Langevin thermostat and Berendsen barostate, NPT-ensemble simulations are performed to generate the phase-trajectory with the time-step as 1 fs, from which the classical mobility of helium in BCC W and Fe could be obtained. The former has been done in Ref.~\cite{wen_many-body_2017}, and the latter is calculated from mean-square displacement
		\begin{equation}
			\mu = \frac{1}{k_\textrm{B}T}\lim_{t\rightarrow\infty}\frac{\left\langle \left[  \bm{R}_\textrm{He}(t) - \bm{R}_\textrm{He}(0)\right] ^2\right\rangle }{6t}
		\end{equation}
	and the many-body dynamical simulations by applied one-dimensional drift-force following, 
		\begin{equation}
			\mu = \lim_{F\rightarrow 0} {\frac{v_d}{F}}
		\end{equation}
	where $\bm{R}_\textrm{He}(t)$ is the instantaneous helium position, $v_d$ is the terminal velocity under the drift force $F$. The detail methodology has been well-documented in Ref.~\cite{wen_many-body_2017}. Fig.~\ref{Fig.5}(b) plots the terminal velocity $v_d$ as function of applied drift force $F$ is plotted in  for helium in BCC W and Fe at 900K, from which the classical mobility is obtained $\mu$= 8.34 and 15.23 $\textrm{\AA}^2$/(ps$\cdot$eV), respectively. It could be seen in Fig.~\ref{Fig.5}(c) and (d), $\mu$ in Fe is almost twice of that in W, both of which reveal complicated non-Arrhenius behaviors at $T$$>$500K. Our results are in good agreement with the data from literature \cite{perez2014diffusion,stewart_atomistic_2011}. The mystery associated with the non-equilibrium nature could be disclosed from the many-body stochastic dynamics of helium diffusion in the following. 
	\begin{table}[tp]  
		\centering  
		\fontsize{8}{8}\selectfont  
		\begin{threeparttable}  
			\caption{Migration energy barrier $\Phi_m$ (in eV) of interstitial helium in BCC W and Fe, respectively, obtained by experimental measurement, \emph{ab initio} calculation, empirical potential. Note that the reference data is obtained in the review paper of Trocellier et al. \cite{trocellier_review_2014}, and the references herein.}  
			\label{tab:performance_comparison}  
			\begin{tabular}{ccccc}  
				\toprule  
				\multirow{1}{*}{}&  
				\multicolumn{1}{c}{Experiment}&\multicolumn{1}{c}{\textit{Ab initio}} &\multicolumn{1}{c}{Empirical}&\multicolumn{1}{c}{Present}\cr    
				\midrule  
				{W}&0.24 - 0.32&0.06&0.05 - 0.16&0.145\cr
				{Fe}&0.06 - 0.08&0.06&0.04 - 0.12&0.091\cr
				\bottomrule  
			\end{tabular}  
		\end{threeparttable}  
	\end{table}  
	
	\subsection{\label{Sec.3.2}Simulations of adiabatic relaxation process}
		
	Similar to the vacancy case, before the relaxation process, the system is initialized by setting the atomic configuration at an excited state with various background temperatures $T^0$. The micro-canonical ensemble is used to mimic an adiabatic process, where the lattice constant is set as the equilibrium one under stress-free condition (See in Table~\ref{tab:2}). The phase-trajectory is collected by solving the equations of motion, from which the temporal evolution of average kinetic energy $T(t)$ is recorded to characterize the relaxation behavior, thus the friction coefficient $\gamma$ and equilibration frequency $\omega_A$ are respectively estimated following Eq.~(\ref{Eq.:Ek_damping}), as listed as in Table~\ref{tab:2}. 

		\begin{table}[!tp]  
			\centering   
			\fontsize{8}{8}\selectfont  
			\begin{threeparttable}  
			\caption{The parameters of helium diffusion in BCC W and Fe at various temperatures calculated in current work, with $a$ the equilibrium lattice constant, $\omega_A$ the equilibration frequency, $\gamma$ the friction coefficient, $m^*$ the effective mass, and $\mu_d$ the thermal drag mobility. Here, 1 a.u. $\approx$ 1.04$\times$$10^{-4}$eV$\cdot$$\textrm{ps}^2/\textrm{\AA}^2$.}  
			\label{tab:2}  
			\begin{tabular}{cccccc}  
				\toprule 
				\multicolumn{6}{c}{\textbf{Helium diffusion in BCC W}}  \cr
				\cmidrule(lr){1-6}  
				$T$ &$a$ &$\omega_A$ &$\gamma$ & $m^*$& $\mu_d$\cr
				[K]& [\AA]& [$\textrm{ps}^{-1}$] & [$\textrm{ps}^{-1}$] & [a.u.] & [$\textrm{\AA}^2$/ps/eV]\\
				\midrule
				300& 3.167 & 28.4 & 12.6&57.6& 13.2\\
				600& 3.172 & 26.2 & 13.6&67.4& 10.5\\
				800& 3.177 & 26.9 & 14.9 &64.1& 10.1\\
				900& 3.180 & 26.7 & 16.0 &65.1& 9.25\\
				1000& 3.183 & 26.1 & 16.4 &67.8&  8.65\\
				1100& 3.186 & 26.2 & 17.1 &67.2& 8.40\\
				1200& 3.189 & 26.5 & 17.3 &66.1& 8.41\\
				1500& 3.199 & 26.4 & 17.6 &66.4& 8.27 \\	
				1800& 3.208 & 26.4 & 17.7 &66.5& 8.19\\
				2400& 3.227 & 26.8 & 18.1 &64.6& 8.26 \\
				\bottomrule  
				\multicolumn{6}{c}{\textbf{Helium diffusion in BCC Fe}}  \cr
				\midrule
				100& 2.898 & 39.9 & 13.7&37.6& 18.7\\
				300& 2.902 & 41.1 & 14.4&35.5& 18.8\\
				400& 2.905 & 38.6 & 14.9 &40.2& 16.1\\
				500& 2.907 & 39.8 & 15.3 &37.7& 16.7\\
				600& 2.909 & 40.5 & 15.2 &36.4&  17.5\\
				800& 2.913 & 43.1 & 16.3 &32.2& 18.4\\
				900& 2.915 & 41.1 & 17.3 &35.5& 15.7\\
				1000& 2.918 & 38.1 & 17.3 &41.3& 13.5 \\	
				1100& 2.920 & 36.2 & 17.3 &45.6& 12.2\\
				1200& 2.923 & 39.4 & 17.5 &38.6& 14.3 \\
				\bottomrule  
			\end{tabular}  
			\end{threeparttable}  
		\end{table}
	
		\begin{figure}
			\graphicspath{}
			\makeatletter
			\def\@captype{figure}
			\makeatother
			\centering
			\includegraphics[width=0.48\textwidth]{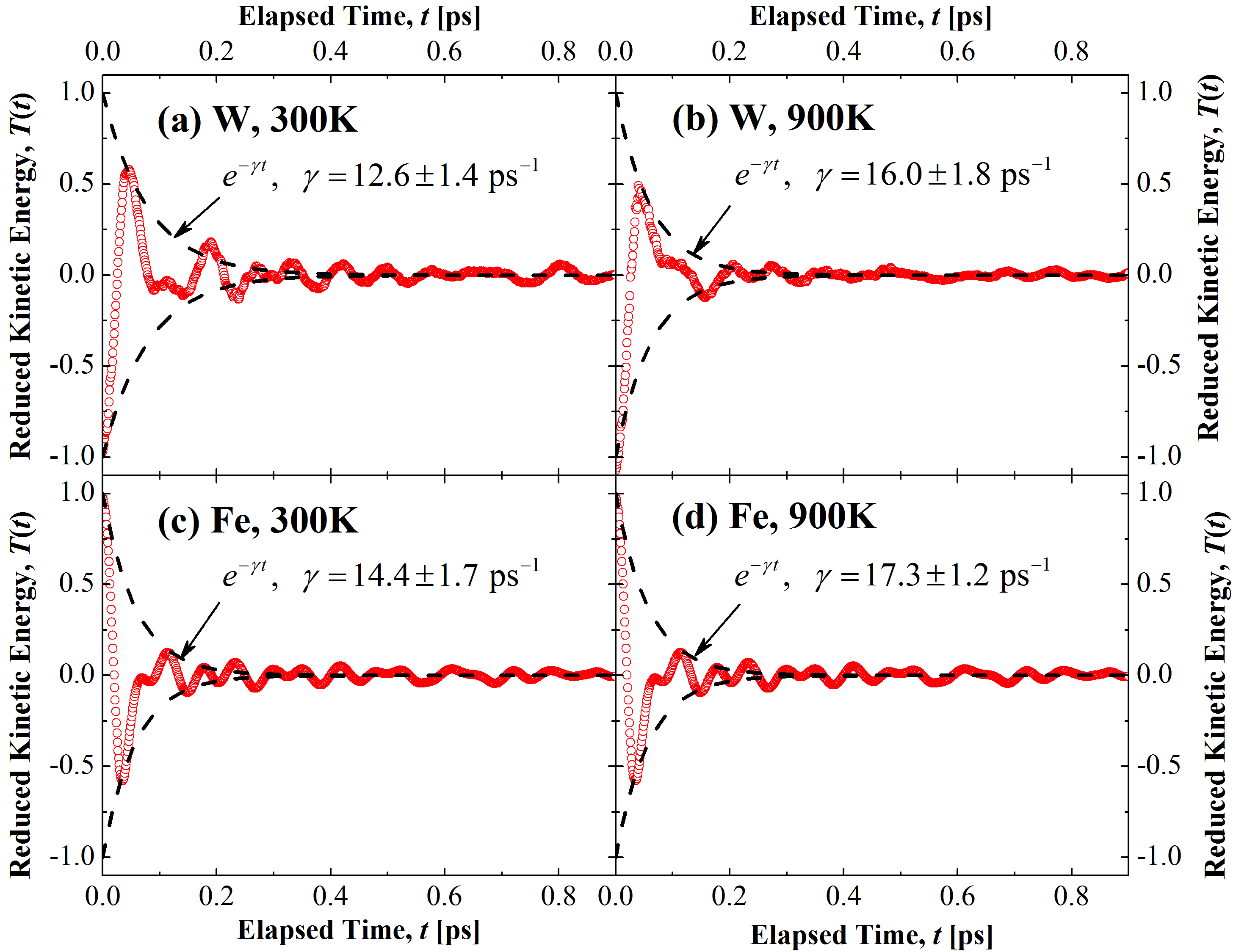}
			\caption{The reduced kinetic energy $T(t)$ of an adiabatic relaxation process for helium migration in BCC W and Fe at 300 and 900K, respectively.}
			\label{Fig.6}
		\end{figure}
				
\begin{figure}
			\graphicspath{}
			\makeatletter
			\def\@captype{figure}
			\makeatother
			\centering
			\includegraphics[width=0.48\textwidth]{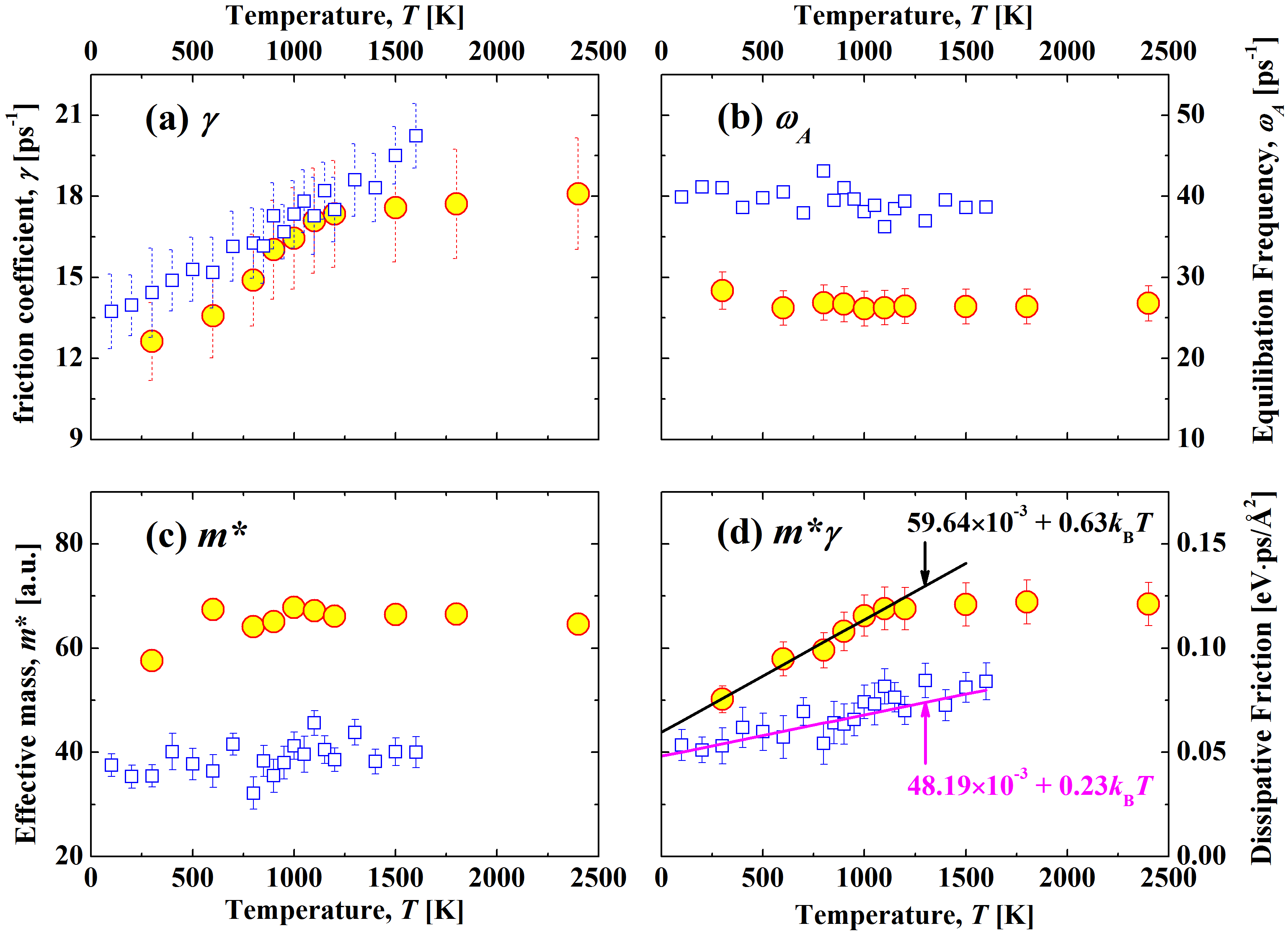}
			\caption{The parameters obtained from adiabatic relaxation process in BCC W (red open circles) and Fe (blue open squares) at various temperatures: (a) friction coefficient $\gamma$; (b) equilibration frequency $\omega_A$; (c) effective mass $m^*$; and (d) dissipative friction $m^*\gamma$, the inverse of classical mobility.}
			\label{Fig.7}
		\end{figure}

	Fig.~\ref{Fig.6} plot the reduced kinetic energy $T(t)$ at 300K and 900K, respectively, of helium in BCC W and Fe. Here, all the curves of $T(t)$ reveal the underdamped behaviors, giving rise to $\gamma/2\omega_A$ should smaller than unity. In addition, plotted in Fig.~\ref{Fig.7}(a), the calculated $\gamma$ of helium in Fe is larger than that in W, but the difference is not significant, both of which reveals increasing temperature dependence, e.g., from $\sim$13.7/ps at 100K to $\sim$20.2/ps at 1600K in Fe, and from $\sim$12.6/ps at 300K to $\sim$17.3/ps at 1200K in W. Here, $\gamma$ for helium diffusion in W are almost fixed as $\sim$18/ps at $T>$1500K, showing a saturated effect of phonon-wind, which is not present $\gamma$ of Fe. Similar behaviors are found in dislocation loop motion \cite{swinburne_classical_2014}, which requires further investigation. In addition, the equilibration frequency $\omega_A$ in both Fe and W are almost temperature-independent, e.g., $\sim$40/ps in Fe and $\sim$27/ps in W, respectively, shown in Fig.~\ref{Fig.7}(b). Correspondingly, the effective mass $m^*$ could be estimated by $m^* = \Phi''(X_A)/g\omega_A^2$, where $g=2/3$ is the geometrical factor for helium diffusion in BCC metals, and $\Phi''(X_A)$ is the curvature of potential $\Phi(X)$ in Fig.~\ref{Fig.5}(a) at equilibrium state $X_A$. $m^*$ plotted in Fig.~\ref{Fig.7}(c) almost keep constant, as $\sim$40 a.u. in Fe and $\sim$65 a.u. in W, respectively, which is far larger than the atomic mass of helium. This is a typical many-body effect. Using the above parameters, the temperature dependence of dissipative friction $m^*\gamma$ is obtained and plotted in Fig.~\ref{Fig.7}(d), as $m^*\gamma$=59.64$\times$$10^{-3}$+0.63$k_\textrm{B}T$ in W at $T$$<$1200K and 48.19$\times$$10^{-3}$+0.23$k_\textrm{B}T$ in Fe, which arises mainly from the temperature dependence of $\gamma$. 
	
	Note that, $\gamma/\omega_A \in [0.4,0.6]$  for helium diffusion in BCC W and Fe at temperatures considered here, leading to the error less than 15\% for the estimation of $\gamma$ (See in \ref{Sec.A2}). As seen in Table.~\ref{tab:2} and Fig.~\ref{Fig.8}, the calculation results of thermal drag mobility $\mu_d$ from adiabatic relaxation process are consistent with the estimated mobility $\mu$ from simulations of many-body dynamics in Sec.~\ref{Sec.3.1} at $T$$>$500K giving rise to $\Phi_m$$\ll$$k_\textrm{B}T$, when the helium diffusion undergoes the Einstein-type non-equilibrium process. Moreover, the difference in $\mu_d$ estimated using simulations of drift-dynamics shown in Fig.~\ref{Fig.5}(b) of helium diffusion in W and Fe is principally the results of difference in effective mass, as well as the equilibration frequency. 
		
	\subsection{\label{Sec.3.3}A coarse-grained formula}
		
	Given an atomistic reaction of hopping between two adjunct potential minima, the classical mobility $\mu$ could be well-defined in the low- and high-temperature limits, respectively, 
		\begin{equation}
			\mu = 
				\left\{
					\begin{aligned}
					& \frac{g\nu_0\lambda^2}{k_\textrm{B}T} e^{-\Phi_m/k_\textrm{B}T}, & \quad \Phi_m\gg k_\textrm{B}T \\
					& \mu_d, & \quad \Phi_m\ll k_\textrm{B}T \\
					\end{aligned}
				\right.
		\end{equation}
	Here, $g=2/3$ is the geometrical factor for helium diffusion in BCC metals. Therefore, the ratio $\kappa = \mu/\mu_d$ should satisfy the following limiting condition as
		\begin{equation}
			\kappa = 
				\left\{
					\begin{aligned}
						& \frac{D_0}{D_E} e^{-1/\mathcal{T}}, & \quad \mathcal{T}\ll 1  \\
						& 1, & \quad \mathcal{T}\gg 1  \\
					\end{aligned}
				\right.
		\end{equation}
	where $D_0 = g\nu_0\lambda^2$ is the pre-factor of diffusivity, $D_E = \mu_d k_\textrm{B}T$ is the Einstein diffusivity, and $\mathcal{T}=k_\textrm{B}T/\Phi_m$ is the effective temperature. Under the prerequisite of meeting two limiting conditions, the problem to propose a coarse-grained formula for $\mu$ at arbitrary $T$ is how to describe the temperature dependence of $\kappa$ for a specific atomistic reaction when $\mathcal{T}$ have a intermediate value. 
	
	Starting from Kramers' theory \cite{KRAMERS1940} on Brownian motion upon a potential force field governed by a generalized Langevin equation (Eq.~\ref{Eq.:Langevin Equation}), substantial progresses have been achieved about the coarse-grained formula for reaction rate. The representative works are respectively the Lifson-Jackson formula \cite{Lifson-Jackson} in the large friction limit
		\begin{equation}\label{Eq.:Lifson-Jackson}
			\kappa = \left[  \int_{0}^{\lambda} e^{\beta \Phi(X)} \textrm{d}X  \int_{0}^{\lambda} e^{-\beta \Phi(X)} \textrm{d}X \right]^{-1}, \quad \gamma \gg \omega_A
		\end{equation}	
	and Risken's expression \cite{Risken1996} based on the cosine-type potential in the low friction limit
		\begin{equation}\label{Eq.:Risken}
			\kappa = \frac{\pi m^*}{\Phi_m}\exp(-\Phi_m/k_\textrm{B}T), \quad \gamma \ll \omega_A
		\end{equation}
	Note that in underdamped limit, the atomic trajectory of point defect shows the long tracks ($\gg$$\lambda$) inside the metals, otherwise the overdamped condition leads to a typical short steps ($\sim$$\lambda$) of hopping between the potential minima \cite{sancho2004diffusion}. The latter is more applicable in mesoscale modeling, e.g., kinetic Monte Carlo simulation. Therefore, we would like to derive our coarse-grained formula on the basis of Eq.~(\ref{Eq.:Lifson-Jackson}) and introduce a tunable parameter to account for the underdamped feature for helium diffusion in metals. 
	\begin{figure}
		\graphicspath{}
		\makeatletter
		\def\@captype{figure}
		\makeatother
		\centering
		\includegraphics[width=0.48\textwidth]{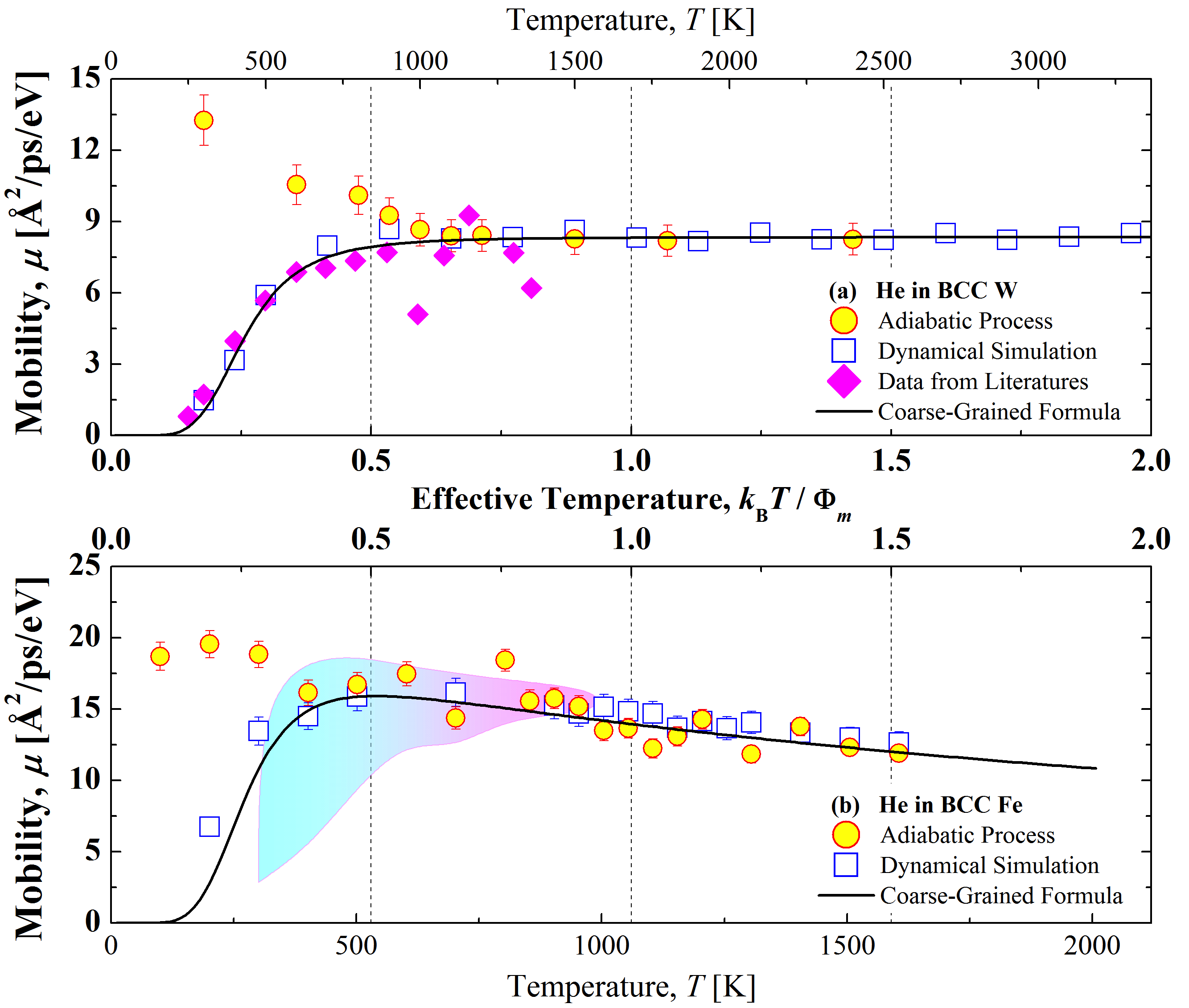}
		\caption{The classical mobility $\mu$ obtained from dynamical simulations in Sec.~\ref{Sec.3.1}, thermal drag mobility $\mu_d$ from adiabatic relaxation process in Sec.~\ref{Sec.3.2} and the data from literature in Fig.~\ref{Fig.5}, as well as the estimation of the coarse-grained formula in Eq.~(\ref{Eq.:coarse-grained}). Here, the shaded area represents the range of other calculations results.}
		\label{Fig.8}
	\end{figure}
	Assuming a cosine-type force field of $\Phi(X)$ exerted by the many-body system as
		\begin{equation}\label{Eq.:cosine_potential}
			\Phi(X) = \frac{\Phi_m}{2}\left[ 1+ \cos(2\pi X/\lambda) \right] 
		\end{equation}
	$\kappa$ is derived following Eq.~(\ref{Eq.:Lifson-Jackson}), as \cite{pavliotis2008diffusive}
		\begin{equation}
			\kappa = J_{0}^{-2}(\frac{1}{2\mathcal{T}}) =
				\left\{
					\begin{aligned}
						& \frac{\pi}{\mathcal{T}} e^{-1/\mathcal{T}}, & \quad \mathcal{T}\ll 1  \\
						& 1, & \quad \mathcal{T}\gg 1  \\
					\end{aligned}
				\right.			
		\end{equation}
	where $J_0(x)$ is the modified Bessel function of the first kind, with $J_0(x)$=1 at $x\ll1$ and $J_0(x)$=$\sqrt{(2\pi x)}e^{x}$ at $x\gg1$, so to meet the high-temperature limit condition. To get the low-temperature limit of $\kappa$, a factor $b$ is defined as
		\begin{equation}
			b \equiv \frac{D_0}{D_E}\frac{\mathcal{T}}{\pi} = \frac{D_0}{\pi\mu_d\Phi_m}
		\end{equation}
	to modified the expression of $\kappa$ as
	\begin{equation}\label{Eq.:modified_kappa}
		\kappa = b \left\lbrace \left[ J_0(\frac{1}{2\mathcal{T}}) - 1\right]^2 + b \right\rbrace^{-1}  
	\end{equation}
	Therefore,  
		\begin{equation}
			\kappa = 
				\left\{
					\begin{aligned}
						& \frac{\pi b}{\mathcal{T}} e^{-1/\mathcal{T}} = \frac{D_0}{D_E} e^{-1/\mathcal{T}}, & \quad \mathcal{T}\ll 1  \\
						& 1, & \quad \mathcal{T}\gg 1  \\
					\end{aligned}
				\right.
		\end{equation}
	satisfying the low- and high-temperature limiting conditions. Finally, a coarse-grained formula for $\mu$ at arbitrary temperature is written as
		\begin{equation}\label{Eq.:coarse-grained}
			\mu = \kappa \mu_d = \frac{b \mu_d}{\left[ J_0(\frac{1}{2\mathcal{T}}) - 1\right]^2 + b}
		\end{equation}
	where the $\gamma^{-1}$ dependence is revealed and consistent with a more rigorous theory proposed in Ref.~\cite{sancho2004diffusion}.
	
		\begin{table}[tp]  
			\centering
			\fontsize{8}{8}\selectfont     
			\begin{threeparttable}  
			\caption{The input parameters of the coarse-grained formula for helium diffusion in BCC W and Fe: $D_0$ the pre-factor, $\Phi_m$ the migration barrier, and $\mu$ the classical mobility.}  
			\label{tab:4}  
			\begin{tabular}{ccccc}  
				\toprule  
				&$D_0$&$\Phi_m$&$\mu_d^{-1}$&$T$\cr
				&[$\textrm{\AA}^2$/ps]&[eV]&[eV$\cdot$ps/$\textrm{\AA}^2$]&[K]\\
				\midrule
				W&3.59& 0.145 & (59.64$\times$$10^{-3}$+0.63$k_\textrm{B}T$), &$<$1200\\
				   &3.59& 0.145 &1/8.34, &$>$1200\\
				Fe&4.45&0.091 & (48.19$\times$$10^{-3}$+0.23$k_\textrm{B}T$)&$>0$\\
				\bottomrule  
			\end{tabular}  
			\end{threeparttable}  
		\end{table}  	 
	Plotted in Fig.~\ref{Fig.8}, by taking the parameters of $\Phi_m$, $\mu_d$ and $D_0$ (listed in Table~\ref{tab:4}) as input, respectively, the predictions of Eq.~(\ref{Eq.:coarse-grained}) applied on the cases of helium migration in BCC W and Fe are in good agreement with the results obtained from many-body dynamics in Sec.~\ref{Sec.3.1} and other calculations. To be honest, the good consistence shown in Fig.~\ref{Fig.8} arises from the appropriate choice of $b$ and modification of $\kappa$ in Eq.~(\ref{Eq.:modified_kappa}). A more precise coarse-grained prediction is obtained from the \emph{diffusion particle} trajectory by solving the generalized Langevin equation. Note that, if using a cosine form of $\Phi(X)$, the effective mass should be directly derived following 
		\begin{equation}
			m_c^* = \frac{\Phi''(X_A)}{g\omega_A^2} = \frac{2\pi^2\Phi_m}{g\omega_A^2\lambda^2}
		\end{equation} 
	with $\Phi''(X_A) = 2\pi^2\Phi_m/\lambda^2$ obtained from Eq.~(\ref{Eq.:cosine_potential}), which is usually different from the prediction $m^*$ from the curvature of potential minima shown in Fig.~\ref{Fig.5}(a) of a parabolic form of $\Phi(X)$, i.e., 
		\begin{equation}
			\Phi(X) = \frac{1}{2}m^*\omega_A^2(X-X_A)^2
		\end{equation}
	 Consequently, $b$ is indeed an effective \emph{quality factor} compatible with the cosine-type potential, as
		\begin{equation}
			b = \frac{D_0}{\pi\mu_d\Phi_m} = \frac{\nu_0\lambda^2}{\pi}\frac{m^*\gamma}{2m_c^*\nu_0^2\lambda^2} =  \frac{m^*}{m_c^*}\frac{\gamma}{\omega_A}
		\end{equation}
	In the ideal case, $m^*/m_c^* = \pi^2/4 \approx 2.47$. 
		
	On the other hand, this coarse-grained formula could be also applied in the mesoscale kinetic Monte Carlo simulation for helium migration, by connecting the reaction rate $\nu$ with the classical mobility $\mu$ following 
		\begin{equation}
			\nu = \frac{\mu k_\textrm{B}T}{g\lambda^2} = \frac{\nu_0 \mathcal{T}}{\pi}\left\lbrace \left[ J_0(\frac{1}{2\mathcal{T}}) - 1\right]^2 + b\right\rbrace ^{-1}
		\end{equation}
	so that $\nu$=$\nu_0$ at $\mathcal{T}$$\ll$1 and $\nu$=$ \mu_d k_\textrm{B}T / \lambda^2$ at $\mathcal{T}$$\gg$1, respectively, satisfies the Arrhenius law and Einstein diffusion theory, where the $\gamma^{-1}$ dependence is again revealed.   
	
\section{\label{Sec.4}Conclusion}
	
	Helium migration is a fundamental atomic activation process of bubble nucleation and growth, affecting the long-term microstructural evolution of structural materials in fission and fusion reactors under irradiation. In order to establish a multi-scale modeling scheme, a practical coarse-grained formula is proposed to describe the temperature dependence of classical mobility of helium migration in BCC W and Fe, where the non-equilibrium nature denoting by the dissipative friction coefficient $\gamma$ is taken into account. Firstly, the dissipation feature of this low-energy atomistic reaction in mesoscale is analyzed, and related it to the microscale many-body dynamics of the thermal excitations involved, so that an analytical expression for $\gamma$ is derived in terms of the dissipation feature of phonon modes by constructing an adiabatic relaxation process. Then, vacancy migration in BCC W is taken as an example and confirms this analytical expression. Further, this adiabatic relaxation simulation method  is applied to get the microdynamic insight of helium diffusion in BCC W and Fe, and get the results in good agreement with experimental data and other calculations. Using the parameters obtained as input, a practical coarse-grained formula, i.e., Eq.~(\ref{Eq.:coarse-grained}), is proposed based on existing formulations for Brownian motion. We expect that current work could help to establish a more universal multi-scale modeling scheme for the microstructural evolution and related phenomena. In our opinion, the current method for the dissipative friction coefficient is not restricted on the specific issue of helium migration in metals, but could be promoted to the atomistic reactions with low-energy pathways in materials science. 
	
\section*{Acknowledgment}	
	The authors would like to express the sincere appreciation of the valuable discussion about this work from Prof. C. H. Woo, Dr. Weijin Chen, Dr. Wenpeng Zhu, and Dr. Long Zhu. This work was supported by the National Key Basic Research Program of China (No. 2015CB351905),  NSFC (No. 11474363, No. 51172291, No. 11602310), to which the authors are thankful. Y. Zheng also thanks support from the Special Program for Applied Research on Super Computation of the NSFC-Guangdong Joint Fund (the second phase), Fok Ying Tung Foundation, Guangdong Natural Science Funds for Distinguished Young Scholar and China Scholarship Council.

\appendix
\section{\label{Sec.A1}{Energy reduction in mesoscale reaction}}
	
\begin{figure}[!t!b]
			\graphicspath{}
			\makeatletter
			\def\@captype{figure}
			\makeatother
			\centering
			\includegraphics[width=0.48\textwidth]{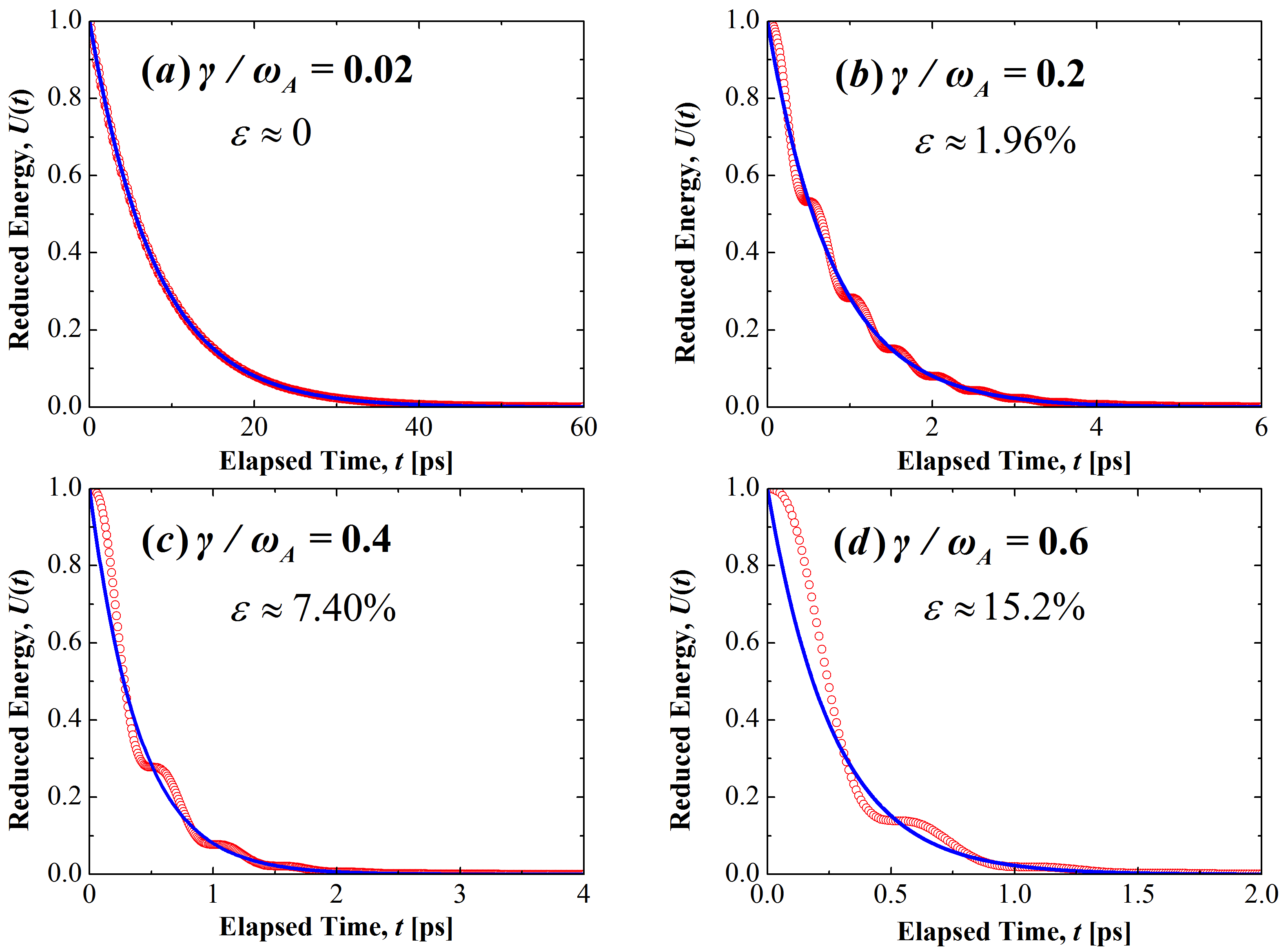}
			\caption{The energy reduction $U(t)$ (red open circles) and its approximate values under $\gamma\ll\omega_A$ when $\gamma/\omega_A$ = 0.02, 0.2, 0.4, and 0.6, respectively, with the corresponding relative error $\varepsilon$ of the estimation of $\tau$ using Eq.~(\ref{Eq.:tau_1}).}\label{Fig.A1}
		\end{figure}

\begin{figure}[!t!b]
			\graphicspath{}
			\makeatletter
			\def\@captype{figure}
			\makeatother
			\centering
			\includegraphics[width=0.48\textwidth]{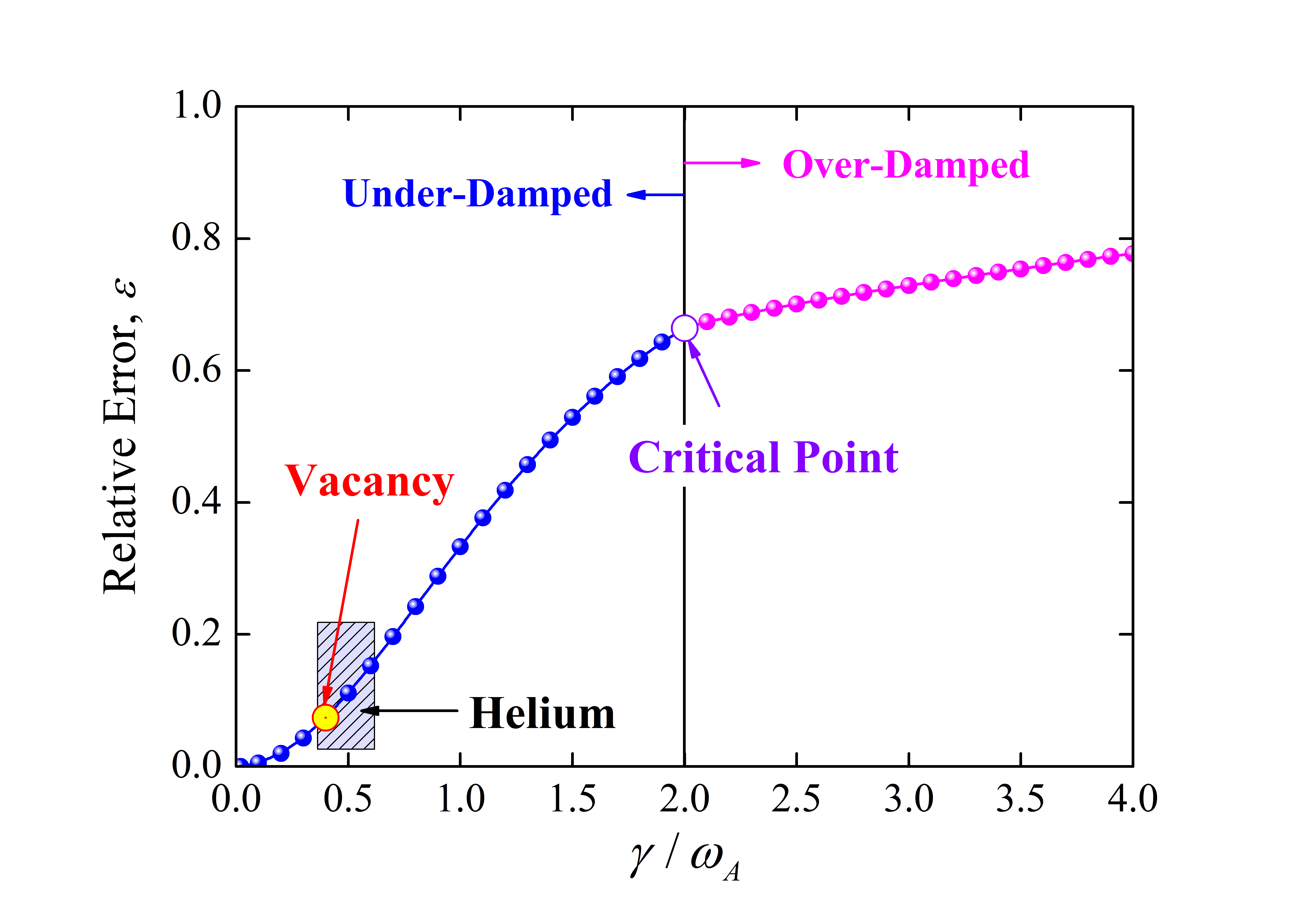}
			\caption{The relative error $\varepsilon$ in the estimation of $\tau$ using the approximation of $\gamma\ll\omega_A$, where $\gamma/\omega_A = 2$ denotes the critical damping. The numerical examples for vacancy and helium migration in current paper correspond to $7.5\% < \varepsilon < 15\%$.}\label{Fig.A2}
		\end{figure}

	Consider an adiabatic relaxation process of phase transport as schematic in Fig.~\ref{Fig.1}, with the equation of motion of phase-space coordinate as in Eq.~(\ref{Eq.:Vac_EOM}). The system is initialized at an excited state $X_0$ near equilibrium with  $\dot{X}(0)$$=$$0$. The initial energy is 
		\begin{equation*}
			U^0 = U(t)|_{t=0} = U(X)|_{X=X_0}  \approx \frac{1}{2} m^* \omega_A^2 X_0^2 
		\end{equation*} 
	Subsequently, without applying random force, i.e., $f(t)=0$ in Eq.~(\ref{Eq.:Vac_EOM}), the system starts to relax adiabatically like an underdamped spring when $\gamma < \omega_A$, with the position and velocity evoking as	
		\begin{equation} \label{Eq.: Vac_Evaluation}
			\left \{
				\begin{aligned}
					X(t) & \approx X_0 e^{- \gamma t/2} \cos{\left( \omega_A t -\frac{\gamma}{2\omega_A} \right)} \\
					\dot{X}(t) & \approx -  \omega_A X(t) \left[ \frac{\gamma}{2\omega_A} + \tan{\left( \omega_A t -\frac{\gamma}{2\omega_A} \right)}\right] \\
				\end{aligned}
			\right.
		\end{equation}	
	which is checked to satisfy the \emph{boundary} conditions		
		\begin{equation} \label{Eq.: Vac_initialcondition}
			\left \{
				\begin{aligned}
					X(t)|_{t=0} & = X_0, \quad & \textrm{and} \quad X(t)|_{t\rightarrow\infty} = 0 \\
					\dot{X}(t)|_{t=0} & = 0, \quad & \textrm{and} \quad \dot{X}(t)|_{t\rightarrow\infty} = 0 \\
				\end{aligned}
			\right.	
		\end{equation}
	Consequently, the evolution of energy $U(t)$ is written as 		
		\begin{equation} \label{Eq.:Ut_macroA}
			\begin{aligned}
				U(t) & = \frac{1}{2} m^* \dot{X}^2(t) + \frac{1}{2} m^* \omega_A^2 X^2(t) \\
					& = \frac{1}{8} m^* \gamma^2 X^2(t) + m^* X^2(t) \frac{\gamma}{2\omega_A} \\
					& \ \ \ \ + \frac{1}{2} m^* X_0^2 e^{- \gamma t} \sin^2{\left( \omega_A t -\frac{\gamma}{2\omega_A} \right)} \\
					& \ \ \ \ + \frac{1}{2} m^* X_0^2 e^{- \gamma t} \cos^2{\left( \omega_A t -\frac{\gamma}{2\omega_A} \right)} \\
					& \approx \frac{1}{2} m^* \omega_A^2 X_0^2 e^{- \gamma t} = U^0 e^{- \gamma t}\\
			\end{aligned}
		\end{equation}	
	Here, we applied the assumption of $\gamma$$\ll$$\omega_A$, and examined it at $\gamma/\omega_A$=0.02, 0.2, 0.4, and 0.6, respectively. As plotted in Fig.~\ref{Fig.A1}, the reduction behavior energy $U(t)$ without approximation is compared to the prediction of Eq.~(\ref{Eq.:Ut_macroA}), and the corresponding relative error $\varepsilon$ of the estimation of $\tau$ using Eq.~(\ref{Eq.:tau_1}). It could be seen from Fig.~\ref{Fig.A1}(a) and (b), the relation of Eq.~(\ref{Eq.:Ut_macroA}) under the approximation of $\gamma\ll\omega_A$ is greatly valid when $\gamma/\omega_A$$<$$0.2$, leading to the relative error $\varepsilon$ of the estimation of $\tau$ following Eq.~(\ref{Eq.:tau_1}) less than 2\%. For the vacancy migration in BCC W (Sec.~\ref{Sec.2}), $\gamma/\omega_A$ is $\sim$0.40, leading to $\varepsilon$$<$10\% shown in Fig.~\ref{Fig.A1}(c), while for helium diffusion in Sec.~\ref{Sec.3.2}, where $\gamma/\omega_A \in [0.4,0.6]$, the relative error is less than 16\%. In this regard, the approximation made in Eq.~(\ref{Eq.:Ut_macroA}) is generally appropriate, as shown as in Fig.~\ref{Fig.A2}.
	

	\section{\label{Sec.A2}{Dynamic response of phonon modes}}
	
	We define the \emph{retarded} Green function $G_r(k\sigma, t)$ and its Fourier transform $G_r(k\sigma,\omega)$ to account for the response of phonon modes under a perturbation, as 	
		\begin{equation}
			G_r(k\sigma,t) = -\textrm{i} \Theta(t) \left\langle [a_{k\sigma}(t), a_{k\sigma}^+] \right\rangle \equiv \langle\langle a_{k\sigma}(t) |  a_{k\sigma}^+ \rangle\rangle
		\end{equation}		
		\begin{equation}
			G_r(k\sigma,\omega) = \int_{-\infty}^{\infty}{G_r(k\sigma,t) e^{-\textrm{i} \omega t} \textrm{d}t}
		\end{equation}	
	where $\Theta(t)$ is the step function. Accordingly, the equation of motion of $G_r(k\sigma,\omega)$ is written as	
		\begin{equation}
			\begin{aligned}
				\hbar \omega \langle\langle a_{k\sigma} |  a_{k\sigma}^+ \rangle\rangle_\omega 
					& = \langle \left[ a_{k\sigma},  a_{k\sigma}^+ \right] \rangle 
					 + \langle\langle \left[ a_{k\sigma}, \mathscr{H}_0 \right] |  a_{k\sigma}^+ \rangle\rangle_\omega \\
					& + \langle\langle \left[ a_{k\sigma}, \mathscr{H}' \right] |  a_{k\sigma}^+ \rangle\rangle_\omega \\
			\end{aligned}
		\end{equation} 		
	Because $\left[ a_{k\sigma},  a_{k\sigma}^+ \right]$$=$$1$ and $\left[ a_{k\sigma}, \mathscr{H}_0 \right]$$=$$\hbar \omega_{k\sigma} a_{k\sigma}$, we have	
		\begin{equation} \label{Eq.: Dyson_0}
			(\omega - \omega_{k\sigma})  \langle\langle a_{k\sigma} |  a_{k\sigma}^+ \rangle\rangle_\omega 	
				= 1 +   \langle\langle \left[ a_{k\sigma}, \mathscr{H}' \right] |  a_{k\sigma}^+ \rangle\rangle_\omega 
		\end{equation}	
	By defining a 'self-energy' $\mathscr{S}_{k\sigma}(\omega)$ to represent the effects of vacancy $\mathscr{H}'$ on phonon modes as 	
		\begin{equation}
			\langle\langle \left[ a_{k\sigma}, \mathscr{H}' \right] |  a_{k\sigma}^+ \rangle\rangle_\omega 
			\equiv \mathscr{S}_{k\sigma}(\omega) \langle\langle a_{k\sigma} |  a_{k\sigma}^+ \rangle\rangle_\omega
		\end{equation}		
	the so-called Dyson equation is obtained, i.e., 	
		\begin{equation}
			\begin{aligned}
				G_r(k\sigma,\omega) & = \langle\langle a_{k\sigma} |  a_{k\sigma}^+ \rangle\rangle_\omega 
					 = \left[ \omega - \omega_{k\sigma} - \mathscr{S}_{k\sigma}(\omega) \right] ^{-1} \\
					& = \left[ \omega - \omega_{k\sigma} - \textrm{Re}\mathscr{S}_{k\sigma} - \textrm{i Im}\mathscr{S}_{k\sigma} \right] ^{-1} \\
			\end{aligned}
		\end{equation}	
	The vibrational frequency $\tilde{\omega}_{k\sigma}$ is thus modified by vacancy with the frequency shift as $\Delta_{k\sigma} \equiv \textrm{Re} \mathscr{S}_{k\sigma}$, the real part of self-energy, as	
		\begin{equation}
			\tilde{\omega}_{k\sigma} = \omega_{k\sigma} + \Delta_{k\sigma} 
		\end{equation}	
	and the image part $\textrm{Im}\mathscr{S}$ corresponds to spectrum width, 	
		\begin{equation}
			\Gamma_{k\sigma} = - \textrm{Im} \mathscr{S}_{k\sigma} (\omega_{k\sigma})
		\end{equation}	
	In this regard, $G_r(k\sigma,\omega)$ can be simplified as	
		\begin{equation}
			G_r(k\sigma,\omega) = \left( \omega - \tilde{\omega}_{k\sigma}+ \textrm{i} \Gamma_{k\sigma} \right)  ^ {-1} 
		\end{equation}		
	Applying the inverse Fourier transform on $G_r(k\sigma,\omega)$, we get	
		\begin{equation} \label{Eq,: G_rt}
			G_r(k\sigma,t) \approx  -\textrm{i} \Theta(t) e^{-\textrm{i} \tilde{\omega}_{k\sigma} t -\Gamma_{k\sigma} t }
		\end{equation}	
	Here, the phonon mode is dissipative due to the presence of vacancy with relaxation time $\tau_{k\sigma}$$=$$\left( 2\Gamma_{k\sigma}\right) ^{-1}$. Accordingly, the energy spectrum $A(k\sigma,\omega)$ is given by	
		\begin{equation}
			\begin{aligned}
				A(k\sigma,\omega)  = - \frac{1}{\pi} \textrm{Im} G_r (k\sigma, \omega) 
					 = \frac{1}{\pi} \frac{\Gamma_{k\sigma}}{\left( \omega - \tilde{\omega}_{k\sigma} \right)^2 + \Gamma_{k\sigma}^2} 
			\end{aligned}
		\end{equation}	
	which is a Lorentzian distribution. In case that the modification of normal modes is usually very tiny, i.e., $\Delta_{k\sigma}$$\ll$$\tilde{\omega}_{k\sigma}$, the relation between $\Delta_{k\sigma}$ and $\Gamma_{k\sigma}$ can be approximately written as \cite{Fultz2010}.	
		\begin{equation} \label{Eq.:Gamma_estimate}
			\Gamma_{k\sigma} \approx \sqrt{2 \tilde{\omega}_{k\sigma} \left| \Delta_{k\sigma}\right| }
		\end{equation}

\section*{Reference}
\bibliographystyle{elsarticle-num}
\bibliography{He_Mig_AM}

\end{document}